\documentclass[nolinenumbers,twocolumn]{aastex701}
 \usepackage{ifpdf}
 \usepackage{graphicx, subfigure}
 \usepackage{natbib}
 \usepackage{times}
 \usepackage{amsmath}
 \usepackage{textcomp}
 \usepackage{lineno}
 \usepackage[colorinlistoftodos,prependcaption,textsize=tiny]{todonotes}
 \usepackage{xcolor}
 \usepackage{gensymb}
\usepackage{comment}

\newcommand{\bbe}{\begin{eqnarray}}
\newcommand{\ee}{\end{eqnarray}}
\newcommand{\n}{\nonumber}

\begin{document}
\nolinenumbers

\title{The Gamma-ray Luminosity Function of Fermi BL Lacertae Objects}

\author[0009-0004-5648-2405]{Garima Rajguru}
\affil{Department of Physics and Astronomy, Clemson University, Kinard Lab of Physics, Clemson, SC 29634-0978, USA}
\email{grajgur@clemson.edu}

\author[0000-0002-8472-3649]{L. Marcotulli}
\affil{Deutsches Elektronen-Synchrotron DESY, Platanenallee 6, 15738 Zeuthen, Germany}
\affil{Department of Physics and Astronomy, Clemson University, Kinard Lab of Physics, Clemson, SC 29634-0978, USA}
\email{lea.marcotulli@desy.de}

\author[0000-0002-6584-1703]{M. Ajello}
\affil{Department of Physics and Astronomy, Clemson University, Kinard Lab of Physics, Clemson, SC 29634-0978, USA}
\email{majello@clemson.edu}

\author[0000-0003-2759-5625]{M. Di Mauro}
\affil{Istituto Nazionale di Fisica Nucleare, Sezione di Torino, Via P. Giuria 1, 10125 Torino, Italy}
\email{dimauro.mattia@gmail.com}

\author[0000-0002-0745-9792]{M. Urry}
\affil{Department of Physics, Yale University, PO Box 208120, New Haven, CT 06520-8120, USA}
\affil{Yale Center for Astronomy and Astrophysics, New Haven, CT 06520, USA}
\email{meg.urry@yale.edu}

\begin{abstract}
\nolinenumbers

We have used the largest sample of gamma-ray-selected \textit{Fermi} BL Lacertae objects (1035 BL Lacs) ever used to determine the luminosity function and its evolution through cosmic time. This sample has its survey biases and selection effects well studied and a robust calculation of its detection efficiency. It spans a large range in luminosity ($L_{\gamma,min}$ = $5.71\times10^{40}$\,erg s$^{-1}$ and $L_{\gamma,max}$ = $1.76\times10^{48}$\,erg s$^{-1}$) and redshift ($ 0 < z \lesssim 4 $). We find that the local luminosity function follows a double power law. The density evolution of BL Lacs depends on the luminosity of the objects, with low-luminosity objects showing negative evolution (i.e., they are less common in the past) and high-luminosity objects showing positive evolution (i.e., they are more common in the past). We confirm that gamma-ray selected BL Lacs and FSRQs evolve similarly depending on their luminosities.

\end{abstract}



\section{Introduction} \label{sec:intro}
The pursuit of the luminosity function (LF) of quasars \citep{Schmidt1968} started soon after the discovery of the first quasar, 3C 273, in 1963 \citep{Schmidt1963}. In the years following their initial discovery, quasars were identified as a powerful subclass of active galactic nuclei (AGNs), with blazars representing AGNs whose jets are oriented along our line of sight \citep{1995Urry}. Blazars show broadband emission from radio waves to $\gamma$-rays and their jets are relativistically beamed towards us \citep[see, e.g.,][]{Padovani2017_Review}. The blazar spectral energy distribution (SED) shows a double-humped structure, with a low-energy hump spanning radio to UV or X-ray energies, attributed to synchrotron emission from electrons in the jet \citep[see, e.g.,][]{FermiBlazarSED2010}. The high-energy hump (keV-GeV) originates from inverse Compton scattering of the jet photons (or external photons from the broad line region, torus, etc.) by the relativistic jet electrons, where the former are upscattered to much higher energies by the latter \citep[see, e.g.,][]{Blandford2019_JetReview}. 

Blazars are further subdivided into flat-spectrum radio quasars (FSRQs), if they show strong emission lines in their optical spectra (equivalent width or EW $>5 \rm\AA$), or BL Lacertae objects (BL Lacs), which do not \citep{Stickel1991, Stocke1991, 1995Urry}. It has been observed that an FSRQ can appear as a BL Lac when the jet emission overpowers the line emission from the broad-line region (BLR) \citep{Carnerero2015,Raiteri2017, Rajagopal2020, Rajguru2024}.

The evolution of blazars is defined as the change in their number and/or luminosity distribution as a function of redshift. The evolution of blazars has been probed across multiple energy bands in the literature: radio waves \citep{Peacock1985, DunlopPeacock1990, PadovaniUrry1992, RectorStocke2001, Wall2005, Padovani2007, Mao2017}, microwave \citep{Giommi2009}, soft X-rays \citep{Padovani1992, GiommiPadovani1994, Rector2000, WolterCelotti2001, Caccianiga2002, Beckmann2003}, hard X-rays \citep{Giommi1991, Ajello2009, Toda2020, Lea2022}, and gamma-rays \citep{Chiang1995, ChiangMukherjee1998, MuckePohl2000, Bhattacharya2009_GLF, Ajello2012, Ajello2014, Broderick2014, Singal2015, Rajguru2025}. Studies of blazar evolution in the gamma-ray regime began by using LF in the X-ray or radio band and extrapolating it based on correlations with gamma-ray luminosities \citep{SteckerSalamon1996, NarumotoTotani2006, InoueTotani2009, SteckerVenters2011}. However, this led to large uncertainties in the models. The advent of the EGRET sample \citep{Hartman1999} provided an opportunity to use gamma-ray data to study the evolution of the blazar directly \citep{Chiang1995, ChiangMukherjee1998, MuckePohl2000, Dermer2007, Bhattacharya2009}. Currently, the Fermi Large Area Telescope \citep[LAT; ][]{LAT_Atwood2009} sample offers the largest sample of gamma-ray-selected blazars and has been used to generate the LF of blazars \citep{Ajello2012, Ajello2014, Rajguru2025}. It has been noted by multiple works that FSRQs evolve positively (i.e., more numerous in the past) until a redshift cutoff that depends on luminosity \citep{DunlopPeacock1990, NarumotoTotani2006, Padovani2007, Ajello2012, Mao2017}. We have also confirmed this in our previous work \citep{Rajguru2025}, and discovered that FSRQs and BL Lacs appear to show similar evolutionary trends in their number density. This can be indicative of a strong genetic link between the blazar subclasses, where BL Lacs and FSRQs could represent the same underlying source, i.e., an accreting black hole, having similar evolution, with differences in their physical properties only.

The evolution of BL Lacs has remained elusive due to small sample sizes, survey biases, and redshift incompleteness. BL Lacs have been shown to follow a variety of cosmological evolution including, positive \citep{Giommi2009, 2013Marcha}, negative \citep{Rector2000, Beckmann2003}, and no evolution \citep{Padovani2007, Bhattacharya2009_GLF}. Characterizing the intrinsic evolution of BL Lacs depends on a large sample that can distinguish among different evolutionary models and redshift estimates available for most of the sources.

A complete understanding of the LF and evolution of blazars is important to address major open-ended questions. The formation and growth of supermassive black holes (SMBHs; $M_{\rm BH} > 10^8 M_{\odot}$) in the early universe is a mystery. Blazars, which host SMBHs at their centers, trace the evolution of SMBHs through cosmic time \citep{Volonteri2011, Sbarrato2015_SMBH, BlazarEpochReionize2024}. The extragalactic gamma-ray background (EGB; \citet{Fichtel1975}) consists of the $\gamma-$ray emission from resolved sources, unresolved sources and diffused processes. Knowing the blazar LF in gamma-rays will allow us to estimate the contribution of unresolved blazars to the EGB correctly \citep{2015Ajello, Lea2020}. An accurate description of blazars' contribution to the EGB will help us understand the contributions of other sources and processes to the EGB.

In this work, we utilize the largest dataset of BL Lacs (1035 sources) ever used to study the  LF at $\gamma$-rays  from 100 MeV to 1 TeV. This has been enabled by the uniform sky survey by \textit{Fermi}-LAT and multiple campaigns to determine the redshift of these sources. This paper is organized in the following manner. In Section \ref{sec:sample}, we discuss the target selection and the data, followed by Section \ref{sec:analysis}, where we elaborate on the analysis technique. The results of our analysis and the corresponding discussions are provided in sections \ref{sec:results} and \ref{sec:conc}. We conclude the paper by summarizing the main implications in section \ref{sec:summary}. Throughout this paper, we use a flat $\Lambda$CDM cosmological model with $H_0=67 \rm ~km~s^{-1}~Mpc^{-1}$, $\Omega_m = 0.3$ and $\Omega_{\Lambda}=0.7$.

\section{Sample} \label{sec:sample}

\begin{table}
\begin{center}
\caption{Composition of the Gamma-Ray Blazar Sample \\ ($|b|> 20^\circ$).}
\begin{tabular}{lcr}
\hline
\hline
Class  &  No. of Objects &Available Redshifts$^a$\\ 
\hline
\hline
Total & 2680 & 1365 \\
\tableline
\tableline
BL Lacs & 1035 & 705 \\
FSRQs & 519 & 519\\
BCUs & 561 & 102\\
Others$^{b}$ & 41 & 39\\
Unassociated& 524 & 0\\
\tableline
\tableline
Redshift Estimate Available  &  1365 & \\
Redshift Estimate Unavailable  &  1315 & \\
\hline
\hline
\multicolumn{3}{l}{
\begin{minipage}{\linewidth} 
\tablenotetext{a}{or spectra with identified features.}
\tablenotetext{b}{Other objects include radio galaxies, narrow-line Seyfert 1 objects, AGNs, etc.}
\end{minipage}
}
\end{tabular}
\label{Tab:sample}
\end{center}
\end{table}

We rely on the sample of \citet{Lea2020} of 2680 sources, of which 1035 are BL Lacs. The catalog comprises sources detected above the Galactic latitude of $|b|> 20^\circ$ and in the energy range of 100 MeV to 1 TeV. \citet{Lea2020} derived the intrinsic source-count distribution associated with this catalog down to $\sim 10^{-10} \rm \, ph\, cm^{-2}\, s^{-1}$ ($\sim 10^{-12} \rm \,  erg\, cm^{-2}\, s^{-1}$), which is an order of magnitude lower than any previous measurement, enabling us to probe lower luminosities than earlier studies. \citet{Lea2020} gives the most accurate calculation of efficiency to date for a Fermi-LAT sample since (1) they used robust Monte Carlo simulations to determine the survey efficiency, taking survey and detection biases into account, and (2) they accounted for any possible curvature in the spectrum of each blazar.

We crossmatch the \citet{Lea2020} sample with the Fermi fourth LAT AGN catalog - Data Release 3 \citep[4LAC-DR3; ][]{4lac_dr3} to obtain the redshifts of the blazars, which are essential in our study, using the Bayesian tool NWAY\footnote{https://github.com/JohannesBuchner/nway} \citep{Salvato_2018_nway}. NWAY provides the probability of any possible counterpart for a given source ($p_{any}$) and the relative posterior probabilities of the potential associations ($p_i$), based on the positions and positional uncertainties of the sources in these two catalogs. For our study, we keep sources with an associated counterpart according to the requirements: $p_{any}>95\%$ and $p_i>95\%$. We find 2156 4LAC-DR3 associations with the sample of \citet{Lea2020}.

Our final catalog consists of 1035 BL Lacs, which is the largest sample used to date to generate the LF  of BL Lacs. The latest data release of 4LAC \citep[4LAC-DR3; ][]{4lac_dr3} and the information available on the website ZBLLAC - A Spectroscopic Library of BL Lac Objects\footnote{\href{https://web.oapd.inaf.it/zbllac/index.html}{https://web.oapd.inaf.it/zbllac/index.html}} were utilized to update the source classifications and redshifts of these objects. Redshifts determined from recent optical spectroscopy studies \citep{Goldoni2021, Redshift_OptSpec_SouthHem} and photometric observations \citep{kaur2017,Yong2024,Sheng2026} were also included in our catalog. 705 of the 1035 BL Lacs have measured redshifts, and the highest redshift in our sample is $z=3.53$. The entire sample is given in Appendix \ref{sec:appx_sample}. Moreover, we have been able to probe three orders of magnitude lower in luminosity than previous studies, with the luminosities in our sample ranging from $L_{\gamma,min}$ = $5.71\times10^{40}$\,erg s$^{-1}$ to $L_{\gamma,max}$ = $1.76\times10^{48}$\,erg s$^{-1}$.

\begin{figure}[htb!]
\centering
\includegraphics[width=\columnwidth]{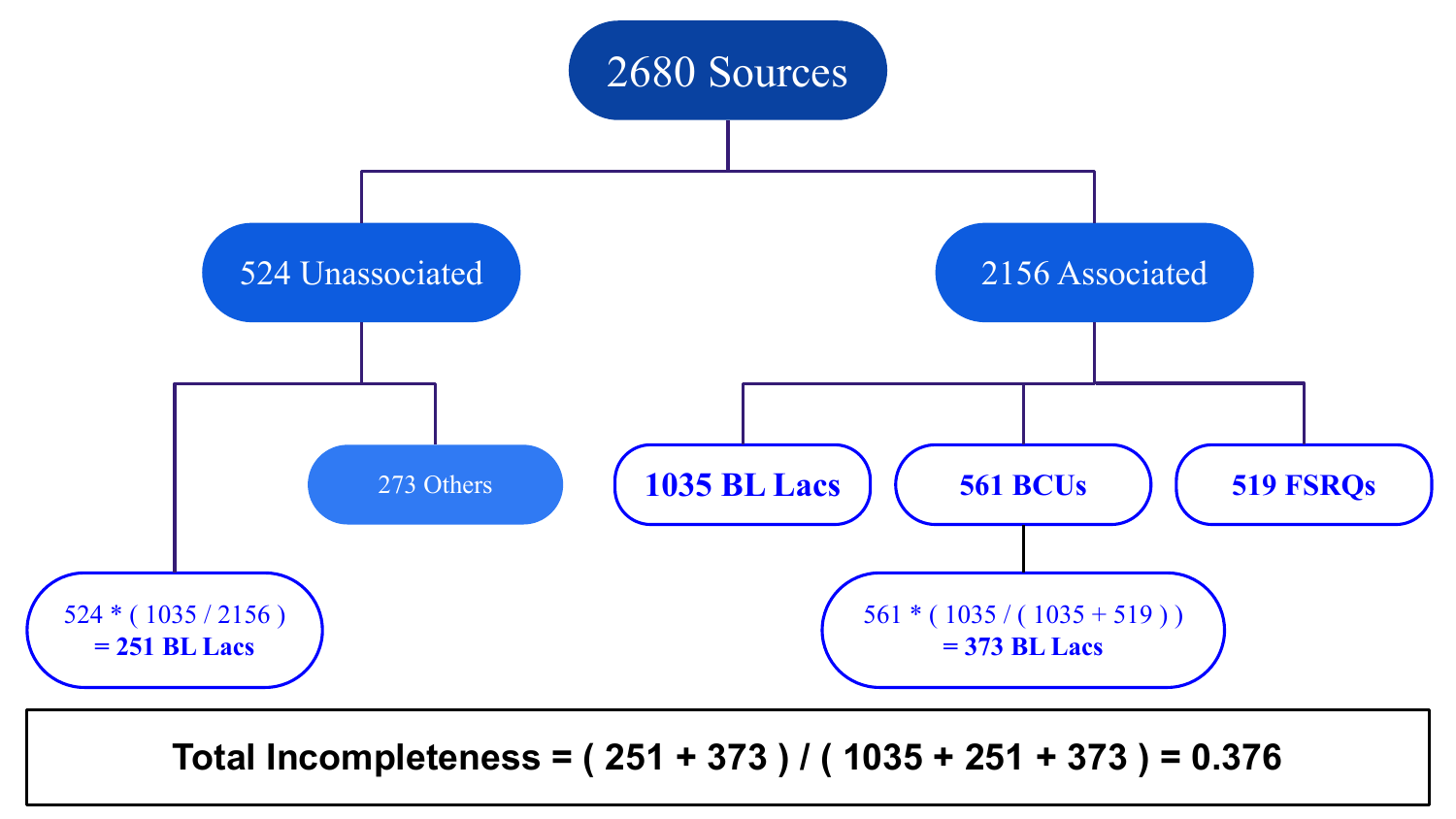}
\caption{The calculation of the incompleteness in our sample is shown. Table \ref{Tab:sample} shows the number of objects in each class. The number of blazars is given in the white rounded rectangles in the diagram. To calculate the incompleteness, we need to account for unclassified BL Lacs. We consider there to be a similar ratio of BL Lacs in the unassociated sources as there are BL Lacs in the total classified population (i.e., 251 BL Lacs in the unassociated sources). Likewise, we consider a similar ratio of BL Lacs in the BCUs as there are BL Lacs in the classified blazars (i.e., 373 BL Lacs in the BCUs). The calculations are shown in the bottom-most rounded rectangles.}
\label{fig:SourceIncompleteness}
\end{figure}

Table \ref{Tab:sample} reports the composition of our sample, 1035 of the 2680 sources are BL Lacs, 519 are FSRQs, and 561 sources are blazar candidates of uncertain type (BCUs). 
The fraction of objects that are unclassified but could be BL Lacs relative to the total sample defines the incompleteness of the sample. Figure \ref{fig:SourceIncompleteness} shows the calculation of the incompleteness in our sample, due to BL Lacs in unassociated sources, as well as among BCUs. We infer the presence of 251 BL Lacs in the unassociated sources and 373 BL Lacs among the BCUs. We multiply the efficiency (see Sec. \ref{sec:ML Fit}) by the completeness of the sample (completeness $=1-0.376 = 0.624$) to account for all the missing potential BL Lacs among the unassociated sources and the BCUs.

\section{Analysis} \label{sec:analysis}

\subsection{Form of the LF} \label{sec:LF}
Our main objective is to simultaneously fit a mathematical expression that describes the blazar LF and its evolution. The LF of a source class is given by the number of sources ($N$) per unit comoving volume ($dV_c$) and luminosity interval ($dL_\gamma$):
\bbe
\label{eq:1}
\Phi_L(L_{\gamma},z) = \frac{d^2N}{dL_{\gamma} dV_c} ,
\ee 
We define the intrinsic joint population density as $\Psi(L_{\gamma}, z, \Gamma)$. To account for the distribution of $\gamma$-ray photon indices ($\Gamma$), we introduce a conditional probability density $p(\Gamma|L_\gamma, z)$, as:
\bbe
\label{eq:2}
\Psi_L(L_{\gamma}, z, \Gamma) = \Phi_L(L_{\gamma}, z) \times p(\Gamma|L_\gamma, z) 
\ee
In Eqn. \ref{eq:2}, $\Phi_L(L_{\gamma}, z)$ is the LF (Eqn. \ref{eq:1}), $L_\gamma$ is the intrinsic rest-frame luminosity between 100 MeV to 1 TeV corrected for EBL absorption  (see Section \ref{sec:EBL}).
Several studies have suggested a luminosity dependence of the photon index \citep{Bhattacharya2009_GLF, Ghisellini2009_BlazarDiv, Meyer2012_PLIndexLum, Ajello2014, Rajguru2025}. In our study, we consider the photon-index probability distribution to be dependent on luminosity but not explicitly dependent on $z$, i.e.,
\bbe
p(\Gamma|L_\gamma,z) &=& p(\Gamma|L_\gamma) \n\\
\Psi_L(L_{\gamma}, z, \Gamma) &=& \Phi_L(L_{\gamma}, z) \times p(\Gamma|L_\gamma)
\ee

We note that the convention used in many multiwavelength LF studies , such as X-ray \citep[see e.g.,][]{Hasinger2005} and radio \citep[see e.g.,][]{Rigby2011} include,
\bbe
\Phi_{\log L} (L_\gamma, z)=\frac{d^2N}{dV_c d\log_{10}L_\gamma} \n 
\ee
We describe the local LF as prescribed in \citet{Ajello2012}, consistent with our previous work in \citet{Rajguru2025}, as:
\bbe
\Phi_L(L_{\gamma}, z=0) = \frac{dN}{dL_\gamma} = \frac{dN}{d\log_{10}L_\gamma}\frac{d\log_{10}L_\gamma}{dL_\gamma} \n\\
= \frac{dN}{d\log_{10}L_\gamma}\frac{1}{\ln(10) L_\gamma} \n \\
\Phi_L(L_{\gamma}, z=0) = \frac{1}{\ln(10) L_\gamma} \Phi_{\log L}(L_{\gamma}, z=0) \n 
\ee
\bbe
\label{eqn:Jacobian}
\Phi_{\log L}(L_{\gamma}, z=0) = \ln(10) L_\gamma \Phi_L(L_{\gamma}, z=0)
\ee
Equation \ref{eqn:Jacobian} gives the Jacobian factor which can be used to convert from one formulation to the other.

Furthermore, for ease of numerical computation, we adopt a reference luminosity $L_0 = 10^{48} \rm \, erg s^{-1}$ and define the dimensionless scaled luminosity $\ell_\gamma$, such that
\bbe
\label{eqn:rescaleL}
\ell_\gamma \equiv \frac{L_\gamma}{L_0} \\
\Phi_\ell(\ell_\gamma,z) = \frac{d^2N}{dV_c d\ell_\gamma} \n 
\ee
\bbe
\label{eqn:rescalePhi}
\Phi(\ell_\gamma,z) \equiv \Phi_\ell(\ell_\gamma,z) = L_0\Phi_L(L_\gamma,z) \\
\Psi(\ell_\gamma,z, \Gamma) \equiv \Psi_\ell(\ell_\gamma,z, \Gamma) = L_0\Psi_L(L_\gamma,z, \Gamma) \n 
\ee
where both $\Phi(\ell_\gamma,z)$ and $\Psi(\ell_\gamma,z, \Gamma)$ have units $\rm Mpc^{-3}$ and $d\ell_\gamma = dL_\gamma/L_0$.
Using Equations \ref{eqn:rescaleL} and \ref{eqn:rescalePhi} in Equation \ref{eqn:Jacobian}, we find that conversion to the logarithmic system is,
\bbe
\Phi_{\log L}(L_{\gamma}, z=0) = \ln(10) \ell_\gamma \Phi(\ell_{\gamma}, z=0) 
\ee

The intrinsic photon index probability distribution is assumed to be a truncated Gaussian function between $\Gamma = 1$ and $\Gamma = 4$, with a luminosity-dependent mean $\mu(\ell_\gamma)$ and dispersion $\sigma$:
\bbe
\label{eq:3}
p(\Gamma|\ell_\gamma) = \frac{1}{C_\Gamma(\ell_\gamma)} e^{-\frac{ (\Gamma-\mu(\ell_\gamma))^2}{2\sigma^2}} ,
\ee
\bbe
\label{eq:4}
\mu(\ell_\gamma) = \mu^* + \beta \times (\log_{10} (L_\gamma ) - 46) .
\ee
The normalization $C_\Gamma(\ell_\gamma)$ takes into account the truncated Gaussian distribution such that:
\bbe
\int_1^4 p(\Gamma\mid \ell_\gamma) d\Gamma=1. \n
\ee

\citet{UrryShafer1984} demonstrated that beam-dominated sources have an LF following a double power-law shape, due to the flattening of the slope of the LF at lower luminosities. Multiple works showed that the blazar LF followed a double power law \citep{NarumotoTotani2006, Ajello2012, 2013Marcha, Ajello2014, Mao2017, Rajguru2025}. Considering these analyses, we describe the local LF at $z=0$ using a smoothly joined double power law \citep[see e.g.,][]{Ueda2003, Hasinger2005, Ajello2014, Rajguru2025}, combined with the photon index distribution.
\bbe
\label{eq:5}
\Psi(\ell_{\gamma}, z=0,\Gamma) =  \n\\
\frac{A}{\ln(10)\ell_{\gamma}} \left[\left(\frac{\ell_{\gamma}}{\ell_{*}}\right)^{\gamma_1}+ \left(\frac{\ell_{\gamma}}{\ell_{*}}\right)^{\gamma2} \right]^{-1}
&\times p(\Gamma|\ell_\gamma)  ,
\ee
where $A$ is the normalization factor, $\ell_{*}$ is the knee of the LF, $\gamma_1$ and $\gamma_2$ are the power-law indices of the low and high-end luminosities.

\subsection{Parametrization of Evolution} \label{sec:Evol}

We use a compound function consisting of the local LF (at $z=0$) and an evolutionary factor ($e(z, \ell_\gamma)$) to study the evolution of BL Lacs \citep[see e.g.,][]{Wall2008, Ajello2014, Lea2022, Rajguru2025}. The evolutionary factor ($e(z, \ell_\gamma)$) depends on redshift. Following our recent work on the FSRQs of the same catalog \citep{Rajguru2025}, we test three different functions to constrain the evolution:
\begin{itemize}
    \item Primarily Density Evolution (PDE), where there is an increase or decrease of the number of sources with redshift. It is quantitatively described by:
    \bbe
    \label{eq:6}
    \Psi(\ell_{\gamma},z,\Gamma) &= \Psi(\ell_{\gamma},z=0,\Gamma) . \n\\
&\times e(z,\ell_\gamma)
    \ee
    \item Primarily Luminosity Evolution (PLE), where there is an increase or decrease in the luminosity of sources with redshift. It is quantitatively described by:
    \bbe
    \label{eq:7}
    \Psi(\ell_{\gamma},z,\Gamma) = \Psi(\ell_{\gamma}/e(z,\ell_\gamma),z=0,\Gamma) .
    \ee
    \item Luminosity-Dependent Density Evolution (LDDE), where there is an increase or decrease in the number of sources which additionally depends on luminosity, as we look into higher redshift. It is quantitatively described by:
    \bbe
    \label{eq:8}
    \Psi(\ell_{\gamma},z,\Gamma) &= \Psi(\ell_{\gamma},z=0,\Gamma) . \n\\
&\times e'(z,\ell_\gamma)
    \ee
\end{itemize}

The evolutionary factors are given by:
\bbe
\label{eq:9}
e(z,\ell_\gamma) = (1+z)^k e^{z/\xi} , \\
k = k^* + \tau \times (\log_{10} (L_\gamma ) - 46) ,\n
\ee
where $z$ is the redshift, $k$ is the luminosity-dependent redshift index, $\tau$ is the parameter that relates the redshift index with the luminosity, and $\xi$ the evolutionary cutoff term; and
\bbe
\label{eq:10}
e'(z,\ell_\gamma) = \left[\left( \frac{1+z}{1+z_c(L_{\gamma})}\right)^{-p1(\ell_{\gamma})} + \left( \frac{1+z}{1+z_c(\ell_{\gamma})}\right)^{-p2} \right]^{-1} ,
\ee
\bbe
\label{eq:alpha}
z_c(\ell_{\gamma}) = z_c^*\cdot (\ell_{\gamma})^{\alpha} , \\
p1(\ell_{\gamma}) =  \delta \times (\log_{10} (L_\gamma ) - 46) . \n
\ee

Here, $z_c^*$ is the redshift peak for a BL Lac with a luminosity of $10^{48}$ erg/s and $z_c(\ell_{\gamma})$ corresponds to the (luminosity-dependent) redshift where the evolution transitions from positive to negative. The parameter $\alpha$ is the power law index connecting the redshift peak of evolution and the luminosity. The evolutionary parameter is dependent on redshift, which depends on indices $p1$ and $p2$. The parameter $\delta$ relates the power-law index to the luminosity. The LDDE model has 10 free parameters, whereas PLE and PDE each have 9. We note that all three models (including PDE and PLE) have a luminosity-dependent density evolution implicit in the evolutionary factor $e(z,\ell_\gamma)$.

\subsection{Maximum Likelihood Fit} 
\label{sec:ML Fit}
We employ the maximum likelihood (ML) approach, introduced by \citet{Marshall1983} and used in \cite{Ajello2009, Ajello2014, Rajguru2025}, to determine the best-fit LF of BL Lacs, given our large sample of BL Lacs spanning large ranges in $L_\gamma$ and $z$. In the literature, the $1/V_{\rm max}$ method given by \citet{Schmidt1968} has been widely used to measure LFs. However, the $1/V_{\rm max}$ method relies on redshift binning, which can bias the results if there is considerable evolution within the bins, especially when the sample spans a large range in luminosity and redshift. Therefore, we have not used this method to generate the LF in our study. \cite{Lynden-Bell_C_1971} developed an unbinned non-parametric maximum likelihood method called the $C$-method, to construct LFs for truncated data. In our study, we have used the parametric unbinned maximum likelihood method introduced by \citet{Marshall1983}, which is best suited for our case. First, this is a parametric approach that utilizes an analytical functional form to fit the data. The best-fit parameters of the function can then be used to further compute various physical quantities of interest. Second, being an unbinned method, it does not depend on binning intervals. Third, it simultaneously fits the local LF and evolution parameters, enabling us to get a comprehensive understanding of LF evolution through a mathematical expression. This is important for blazars, where the LF evolves with both luminosity and redshift, and our fitting procedure accounts for this.

In the ML approach, the luminosity, redshift and photon index space is finely divided into small intervals of size $d\ell_{\gamma}dzd\Gamma$, such that either 0 or 1 object can occupy each interval. The best-fit LF is obtained by comparing, through an ML estimator, the number of observed objects found in a specific interval to the number of expected objects (for a given LF model). The expected number of sources in each interval is given by:
\bbe
\label{eq:11}
\lambda(\ell_{\gamma},z, \Gamma)d\ell_{\gamma}dz d\Gamma   = \Psi(\ell_{\gamma},z, \Gamma) \Omega(\ell_{\gamma},z,\Gamma) \n\\
\times \frac{dV_c}{dz} d\ell_{\gamma} dz d\Gamma
\ee
where $\Omega(\ell_{\gamma},z,\Gamma)$ is the effective sky coverage (for details see Sec. \ref{sec:SkyCoverage}). The effective sky coverage is the effective solid angle over which a source with luminosity $\ell_\gamma$, redshift $z$, and photon index $\Gamma$ can be detected in the LAT survey used in this work. $\frac{dV_c}{dz}$ ($=\frac{dV_C}{dzd\Omega'}$) is the comoving volume element per unit redshift ($z$) and per unit solid angle ($\Omega'$), computed using the prescription given in \citet{Hogg1999}.

The likelihood function ($L$) is based on joint Poisson probabilities for the detection of 1 source in the $i^{th}$ interval and 0 sources in all other intervals:
\begin{align}
\label{eq:12}
L = & \prod_i \lambda(\ell_{\gamma,i},z_i,\Gamma_i) d\ell_{\gamma} dz  d\Gamma e^{-\lambda(\ell_{\gamma,i},z_i,\Gamma_i) d\ell_{\gamma} dz d\Gamma} \n\\
& \times \prod_j e^{-\lambda(\ell_{\gamma,j},z_j,\Gamma_j) d\ell_{\gamma} dz d\Gamma}
\end{align}
For ease of computation, we transform the above into the standard expression $S=-2\ln\ L$ and drop the model-independent terms:
\begin{align}
S = & -2 \sum_i \ln \left( \frac{d^3 N}{d\ell_{\gamma} \, dz \, d\Gamma} \right) \n\\
    & + 2 \int_{\Gamma_{\min}}^{\Gamma_{\max}} \int_{\ell_{\gamma,\min}}^{\ell_{\gamma,\max}} 
    \int_{z_{\min}}^{z_{\max}} \lambda(\ell_{\gamma}, \Gamma, z) \, dz \, d\ell_{\gamma} \, d\Gamma
\label{eq:13}
\end{align}
Unless specified otherwise, the integration limits are $z_{min}$ = 0.0001, $z_{max}$ = 5.0, $\Gamma_{min}\,=$ 1.0 and $\Gamma_{max}\,=$ 4.0. $\ell_{\gamma,min}$ and $\ell_{\gamma,max}$ correspond to the minimum and maximum scaled luminosity of the sample, respectively, which is different for the different catalogs generated via the redshift extraction method (see section \ref{sec:Redshifts}), with typical values of $\ell_{\gamma,min}$ = $5.71\times10^{40}$\,erg s$^{-1} /L_0$, $\ell_{\gamma,max}$ = $1.76\times10^{48}$\,erg s$^{-1} /L_0$. Due to the power-law shape of the LF, changing the upper limits of the integrations has no appreciable effect. The best-fit parameters have been obtained by minimizing the above using the Python interface for MINUIT2 minimization package, embedded in ROOT\footnote{\url{https://root.cern/}}, called \texttt{iminuit}\footnote{\url{https://pypi.org/project/iminuit/}} \citep{James:1975dr, iminuit}. We use the Akaike Information Criterion \citep[AIC;][]{AIC_1974} to evaluate the goodness of fit. The AIC estimation penalizes each additional degree of freedom in complex models compared to simple models, in addition to the value of $-2\ln L$ when the function is minimized. This is useful when comparing models with different degrees of freedom. The best-fit model is the one with the lowest AIC.

The value of $S$ and its associated 1\,$\sigma$ uncertainty are determined by varying the parameter of interest while allowing all other parameters to float, until a change of $\Delta S = 1$ is reached, leading to an estimate of the 68\,\% confidence interval for the parameter of interest \citep{avni1976}. Despite being computationally intensive, Eq.~\ref{eq:13} offers the advantage of treating each source's $k$-correction individually. The redshift and the photon index of each source allow us to calculate the corresponding $k$-correction, which can then be used to estimate the corrected rest-frame luminosity from the observed flux (see Equation \ref{eqn:K-correction} in Section \ref{sec:EBL} for the calculation of the $k$-correction).

To assess the performance of the best-fit LF, we compare the observed redshift, photon index, intrinsic source count and luminosity distributions of the data to the prediction of the LF. These distributions can be derived directly from the LF, as follows:
\begin{eqnarray}
\frac{dN}{dz} & = & \int^{\Gamma_{max}}_{\Gamma_{min}} \int^{\ell_{\gamma,max}}_{\ell_{\gamma,min}} \lambda(\ell_{\gamma},\Gamma,z) d\ell_{\gamma}  d\Gamma \\
\frac{dN}{d\ell_{\gamma}} & = & \int^{\Gamma_{max}}_{\Gamma_{min}} 
\int^{z_{max}}_{z_{min}} \lambda(\ell_{\gamma},\Gamma,z) dz d\Gamma \\
\frac{dN}{d\Gamma}  & = &  \int^{\ell_{\gamma,max}}_{\ell_{\gamma,min}} 
\int^{z_{max}}_{z_{min}} \lambda(\ell_{\gamma},\Gamma,z) dz d\ell_{\gamma}
\end{eqnarray}
\begin{align}
N(>S) =
\int^{\Gamma_{max}}_{\Gamma_{min}} 
\int^{z_{max}}_{z_{min}}  \int^{\ell_{\gamma,max}}_{\ell_{\gamma}(z,S, \Gamma)}
 \Phi(\ell_{\gamma},z) p(\Gamma|\ell_\gamma) \frac{dV_c}{dz} d\ell_{\gamma} dz d\Gamma
\label{eq:logn}
\end{align}
where $\ell_{\gamma}(z,S, \Gamma)$ is the scaled luminosity of a source at redshift $z$ having a flux of $S$ and photon-index $\Gamma$. 

We employ the ``N$^{obs}$/N$^{mdl}$'' method to display the LF \citep{LaFranca1997, Miyaji2001, LaFranca2005, Hasinger2005, Ajello2014, Rajguru2025}. After obtaining the best-fit LF model, it is possible to determine the value of the model-corrected binned LF estimates in a given luminosity and redshift bin from the intrinsic LF, by:
\begin{equation}
\Phi(\ell_{\gamma,i},z_i) = \Phi^{mdl}(\ell_{\gamma,i},z_i) \frac {N^{obs}_i}{N^{mdl}_i}
\label{eq:Nobs}
\end{equation}
where $\ell_{\gamma,i}$ and $z_i$ are the scaled luminosity and redshift of the i$^{th}$
bin and $\Phi^{mdl}(\ell_{\gamma,i},z_i)$ is the best-fit LF model. $N^{obs}_i / N^{mdl}_i$ is the ratio of the observed and predicted number of BL Lacs in that bin, which is used to scale the LF and obtain its model-corrected value. These model-corrected binned LF estimates primarily serve as a diagnostic of the agreement between the best-fit LF model and the observed source distribution locally in luminosity and redshift. These two approaches, i.e. the maximum likelihood (ML) method of \citet{Marshall1983} along with the ``N$^{\text{obs}}$/N$^{\text{mdl}}$" estimator, offer a minimally biased estimate of the LF \citep[see also][]{Miyaji2001}.

\subsection{Deriving the Intrinsic Luminosity $L_\gamma$ - EBL Correction}
\label{sec:EBL}
The Extragalactic Background Light \citep[EBL; see e.g.][]{Mattila2019} is the integrated emission originating from stars, galaxies and accreting compact objects across the history of the universe from the UV to the infrared range  \citep{Peebles1993, Hauser1998, Dwek2013, Mattila2019, Baxter2026, Banerjee2026_EBL}. High-energy $\gamma$-ray photons can pair produce with EBL photons, leading to an attenuation of the observed $\gamma$-ray flux of a distant source. 
This reduction in flux is non-negligible in the spectra of BL Lacs, which emit photons at $E>10 \,\rm GeV$ energies, at which the EBL attenuation becomes important.

To account for this effect, we utilize the intrinsic blazar spectrum given in \citet{Lea2020}:
\bbe
\label{eqn:spectra}
\frac{dN}{dE} = K\left[\left(\frac{E}{E_b}\right)^{1.7} + \left(\frac{E}{E_b}\right)^{2.8}\right]^{-1} 
\ee
where $K =$ normalization and $E_b = $ the break energy related to the photon-index ($\Gamma$) of each source by the relation $\log E_b (\rm GeV) = 9.25 - 4.11 \Gamma$.

We calculate the intrinsic spectral flux for each BL Lac using its photon index ($\Gamma$) in Equation \ref{eqn:spectra}. Using the redshift of each source, we also calculate the EBL-attenuated spectral flux of each source in the energy range between 100 MeV and 1 TeV, using the EBL model given in \citet{Saldana-Lopez2021} and the optical depths given in \citet{Dominguez2023b}. The correction factor ($C_{\rm EBL}(z,\Gamma)$) is given by the ratio,
\bbe
C_{\rm EBL}(z,\Gamma)=\frac{\displaystyle\int_{E_{\min}}^{E_{\max}}E\frac{dN}{dE}dE}{\displaystyle\int_{E_{\min}}^{E_{\max}}E\frac{dN}{dE}e^{-\tau(E,z)}dE}.
\ee
The intrinsic EBL-corrected flux is then calculated by multiplying the observed flux of the source by $C_{\rm EBL}(z,\Gamma)$:
\bbe
F_{\rm int}=F_{\rm obs}\,C_{\rm EBL}(z,\Gamma)
\ee
The corresponding $k$-correction factor is numerically computed as,
\bbe
\label{eqn:K-correction}
K_\gamma(z,\Gamma)=\frac{\displaystyle\int_{E_{\min}/(1+z)}^{E_{\max}/(1+z)}E\frac{dN}{dE}dE}{\displaystyle\int_{E_{\min}}^{E_{\max}}E\frac{dN}{dE}dE}
\ee
Finally, the rest-frame intrinsic luminosity $L_\gamma$ is calculated using the EBL correction and the $k$-correction factor, using the equation:
\bbe
\label{eqn:luminosity}
L_\gamma=4\pi d_L^2(z)\,F_{\rm obs}\,C_{\rm EBL}(z,\Gamma)\,K_\gamma(z,\Gamma)
\ee
The luminosity $L_\gamma$ is then used to compute the LF of BL Lacs. Throughout this work, unless explicitly stated otherwise, $L_\gamma$ refers to the EBL-corrected, rest-frame intrinsic luminosity defined in Equation \ref{eqn:luminosity}. All subsequent analyses, including the LF modeling and the quantities derived from the LF, are based exclusively on these EBL-corrected luminosities.

As a consistency check, we repeat the LF analysis without applying the EBL correction and present the results in Appendix \ref{sec:NoEBL_LF}. The resulting LF is consistent with our EBL-corrected analysis within the uncertainties, and the main conclusions remain unchanged.

\subsection{Dealing with Missing Redshifts}
\label{sec:Redshifts}
Of the 1035 sources, 705 BL Lacs have redshifts available in the literature, i.e., 31\% of BL Lacs do not have a redshift. \citet{Ajello2014} had used a complete sample of 211 Fermi BL Lacs to generate the intrinsic redshift distribution ($dN/dz$). They included both spectroscopic redshift measurements and photometric constraints to derive redshift estimates for the entire sample and generate an accurate redshift distribution. 
We undertake a Monte Carlo approach using the intrinsic $dN/dz$ of \citet{Ajello2014} to analyze our sample:
\begin{enumerate}
\item We generate 1000 bootstrap samples from the initial 1035 BL Lac sample by randomly drawing sources with replacement from the parent sample \citep[see e.g.,][]{efron1994bootstrap}.
\item  For each source in a given bootstrap realization, we retain the given flux, redshift and photon index of the source. If a source lacks a redshift measurement, we assign a redshift value to it by sampling from the intrinsic $dN/dz$ of \citet{Ajello2014}. This ensures that all BL Lacs in our 1000 catalogs have a redshift.
\item We use the ML routine described in section \ref{sec:ML Fit} to obtain the best-fit LF for each of the 1000 catalogs. We obtain the final LF by taking the median of all the 1000 LFs. The statistical uncertainty is given by the 68\% containment region around the median. We note that the statistical uncertainty arises from the  random scatter due to redshift assignments within the adopted $dN/dz$ prior.
\end{enumerate}

The luminosity $L_\gamma$ vs redshift plot of the BL Lacs in one representative sample among the 1000 samples is shown in Appendix \ref{sec:LzPlot}.

\subsection{The Sky Coverage}
\label{sec:SkyCoverage}
The detection efficiency associated with the sample at $|b|> 20^\circ$, along with its corresponding uncertainty, is reported in \citet{Lea2020}. We denote the  dimensionless, flux-dependent detection efficiency by $\omega(S)$, where $0<\omega (S)<1$. It represents the probability of detecting in this survey a blazar with flux $S$, taking into account the survey biases and selection effects. We define the effective sky coverage as,
\bbe
\Omega(S) = \omega(S) \times  \Omega_{\rm geom} \n
\ee
where $\Omega_{\rm geom}$ is the geometric solid angle of the sky surveyed by the telescope for $|b|> 20^\circ$, which is 27143.61 $\rm deg^2$. The detection efficiency ($\omega$) depends on the flux and, therefore, the luminosity, $z$ and $\Gamma$:
\bbe
\Omega(\ell_\gamma,z,\Gamma) = \Omega(S) = \omega\left[S(\ell_\gamma,z,\Gamma)\right] \times  \Omega_{\rm geom} \n
\ee
Thus, very bright sources can possess a maximum effective sky coverage near the total geometric solid angle of 27143.61 $\rm deg^2$.

\subsection{Systematic Uncertainties}
\label{sec:Systematics}

There are two major avenues of systematic uncertainty in our analysis:
\begin{enumerate}
\item \textit{Uncertainty due to Redshift Extraction:} We study two extreme scenarios of the redshift extraction method in order to estimate the uncertainty related to redshift extraction. First, we restrict the analysis to the 705 BL Lacs with measured redshifts and run the full maximum-likelihood (ML) fitting pipeline on this subsample. The resulting LF is corrected for both type and redshift incompleteness (a factor of 0.57). Note that adopting the 705 BL Lac sample is a rather extreme hypothesis as we know \citep{Ajello2014} that most BL Lacs that do not have a spectroscopic redshift are at higher redshifts ($z>0.5$) on average. 

Second, we construct an expanded catalog of 1035 BL Lacs by assigning redshifts following the procedure described in Section~\ref{sec:Redshifts}, but imposing the constraint $z_{\rm extracted} > 0.5$. The redshift $z = 0.5$ is chosen as a lower limit because the intrinsic $dN/dz$ in \citet{Ajello2014} shows a significant population of BL Lacs above this redshift, most of which were discovered using photometric redshift measurements. We then determine the best-fit LF for this alternative catalog. These two extreme assumptions yield two distinct best-fit LF solutions that are associated with the alternative assumptions about the missing-redshift distribution.
\item \textit{Uncertainty due to Sky Coverage:}  We calculate the upper and lower bounds of the efficiency using the uncertainty associated with the detection efficiency given in \citet{Lea2020} (see Sec.~\ref{sec:SkyCoverage}). The fitting procedure is repeated using these limiting efficiency curves to obtain the corresponding extrema of the model parameters. This analysis is performed separately for each of the two redshift cases described above, resulting in four LF realizations, i.e., the upper and lower bounds for each redshift case.
\end{enumerate}
The overall systematic uncertainty is defined by the envelope spanned by these four LF solutions, with the maximum and minimum values providing the upper and lower bounds, respectively. We note that the uncertainty due to the detection efficiency is negligible compared to the uncertainty due to redshift extraction. The diagnostic plots and the source-count distribution in Figures \ref{fig:ldde_panels} and \ref{fig:dNdS} show only statistical uncertainties, whereas the LF-derived quantities in Figures \ref{fig:LF_Evol_L},  \ref{fig:LF_Evol_z}, \ref{fig:LF_DeEvol}, \ref{fig:LumDens} and \ref{fig:NumDens} include both statistical and systematic uncertainties. For example, in Figure \ref{fig:LF_Evol_L} the light blue shaded region is the systematic uncertainty band. 

\subsection{Variability}
\label{sec:Variability}
Since blazars are variable objects, we consider the effect of variability on our study. First, we have used average quantities throughout our work, such as mean flux, mean luminosity, and mean photon index, to mitigate the effects of variability. 
\citet{Ajello2014} measured the BL Lac LF using only 1 year of data. If source variability were a problem, we would expect a marked difference between that LF and ours (which relies on 8\,yr of data). Instead, we find (see Section~\ref{sec:results}) that most parameters are consistent within the uncertainty limits, indicating that the effect of variability is negligible in our study. Moreover, in \citet{Rajguru2025}, we showed that variability did not adversely affect our LF analysis for FSRQs. Since FSRQs are more variable than BL Lacs, this suggests that the analysis for BL Lacs would not be harmed by variability.

\section{Results} \label{sec:results}
\subsection{Primarily Luminosity Evolution}

\begin{deluxetable*}{cccccccccccc} 
\tablecaption{Best-fit parameters of the Primarily Luminosity Evolution (PLE) LF.}
\label{tab:PLE}
\tablewidth{\textwidth} 
\setlength{\tabcolsep}{3pt}
\tabletypesize{\scriptsize}
\tablehead{
\colhead{$\mu$\textsuperscript{a}} & \colhead{$\sigma$} & \colhead{$\beta$} & \colhead{$A$\textsuperscript{b}} &
\colhead{$\ell_*$} & \colhead{$k^*$} & \colhead{$\gamma_1$} & \colhead{$\gamma_2$} &
\colhead{$\xi$} & \colhead{$\tau$} & \colhead{$-2\log L$} & \colhead{AIC}
}
\startdata
$2.029^{+0.007}_{-0.007}$ &
$0.204^{+0.005}_{-0.005}$ &
$0.006^{+0.005}_{-0.005}$ &
$2.9^{+3.0}_{-1.4}\times 10^5$ &
$0.0047^{+0.0041}_{-0.0026}$ &
$2.03^{+0.45}_{-0.48}$ &
$0.65^{+0.06}_{-0.07}$ &
$1.55^{+0.13}_{-0.12}$ &
$-0.62^{+0.08}_{-0.11}$ &
$0.83^{+0.09}_{-0.09}$ &
$-8453$ &
$-8435$
\enddata
\tablecomments{
\textsuperscript{a} Parameter values reported are the median of the best-fit values obtained for all 1000 Monte Carlo samples. Statistical uncertainties are reported in this table, computed as the 68\% containment region around the median value.  \\
\textsuperscript{b} In units of $10^{-13}$\,Mpc$^{-3}$. \\
}
\end{deluxetable*}

In Table \ref{tab:PLE} we show the best-fit parameters of the PLE model and the uncertainties calculated as the 68\% containment region around the median value of the analysis using 1000 Monte Carlo samples. A positive value of $k$ indicates that BL Lacs were more luminous in the past than they are today (positive evolution). Note that $k^{*}_{BLL} =  2.03^{+0.45}_{-0.48}$ is lower than $k^{*}_{FSRQ} = 3.09 \pm 0.55$ \citep{Rajguru2025}, which means that the positive evolution in luminosity was more prominent in the case of FSRQs than it is for BL Lacs. The cutoff in the evolutionary parameter is $\xi = -0.62^{+0.08}_{-0.11}$, using which we can derive the redshift at which the distribution peaks. Differentiating Eqn. \ref{eq:9}, we get $z_{peak} = -1-k\xi$. For a luminosity $L_\gamma = 10^{46} \rm erg/s$, the redshift peak is at $z_{peak} = 0.25$. This is at a lower value than for FSRQs \citep[$z_{peak, \rm FSRQ} = 0.42$; ][]{Rajguru2025} and BL Lacs in \citet[][$z_{peak, \rm BLL} = 0.8$]{Ajello2014} 
at the same luminosity $L_\gamma = 10^{46} \rm erg/s$. The reduction in $z_{peak}$ inferred in this work can be explained by the fact that we probed three orders of magnitude lower in luminosity than previous studies.

\subsection{Primarily Density Evolution}

\begin{deluxetable*}{llllllllllll}
\tablecaption{Best-fit parameters of the Primarily Density Evolution (PDE) LF.}
\label{tab:PDE}
\tablewidth{0pt}
\setlength{\tabcolsep}{3pt}
\tabletypesize{\scriptsize}
\tablehead{
\colhead{$\mu$\textsuperscript{a}} & \colhead{$\sigma$} & \colhead{$\beta$} & \colhead{$A$\textsuperscript{b}} &
\colhead{$\ell_*$} & \colhead{$k^*$} & \colhead{$\gamma_1$} & \colhead{$\gamma_2$} &
\colhead{$\xi$} & \colhead{$\tau$} & \colhead{$-2\log L$} & \colhead{AIC}
}
\startdata
$2.030^{+0.007}_{-0.007}$ &
$0.204^{+0.005}_{-0.005}$ &
$0.006^{+0.005}_{-0.005}$ &
$3.5^{+3.1}_{-1.6}\times 10^5$ &
$0.0042^{+0.0037}_{-0.0021}$ &
$4.82^{+1.26}_{-1.29}$ &
$0.67^{+0.86}_{-0.07}$ &
$1.49^{+0.14}_{-0.84}$ &
$-0.26^{+0.04}_{-0.05}$ &
$2.05^{+0.31}_{-0.31}$ &
$-8457$ &
$-8439$
\enddata
\tablecomments{
\textsuperscript{a} Parameter values reported are the median of the best-fit values obtained for all 1000 Monte Carlo samples. Statistical uncertainties are reported in this table, computed as the 68\% containment region around the median value. \\
\textsuperscript{b} In units of $10^{-13}$\,Mpc$^{-3}$. \\
}
\end{deluxetable*}

The positive evolution ($k>0$) in the PDE model indicates that BL Lacs were more numerous in the past than they are today. The best-fit parameters of the PDE model are shown in Table \ref{tab:PDE}. For the PDE model, we find that $k^* = 4.82^{+1.26}_{-1.29}$, which is smaller than $k^{*}_{FSRQ} = 11.26 \pm 1.41$ \citep{Rajguru2025}, leading to a similar conclusion that positive evolution in density was more prominent in the case of FSRQs than it is for BL Lacs. We use the values $k^* = 4.82^{+1.26}_{-1.29}$ and $\xi = -0.26^{+0.04}_{-0.05}$ to calculate the redshift at which the number distribution peaks, following the same equation as shown in the PLE case. For a luminosity $L_\gamma = 10^{46} \rm erg/s$, the redshift peak is at $z_{peak} = 0.25$. This agrees with the value of $z_{peak}$ found in the PLE case. Tables \ref{Tab:BestModelInference} show that the value of $-2\ln L$ is lower for the PDE model ($-8457$) than for the PLE model ($-8453$). Similarly, the AIC value of the PDE model is also lower than that of the PLE model.

\subsection{Luminosity-Dependent Density Evolution - Best Fit LF}

\begin{deluxetable*}{lllllllllllll}
\tablecaption{Best-fit parameters of the Luminosity-Dependent Density Evolution (LDDE) LF.}
\label{tab:LDDE}
\tablewidth{\textwidth}
\setlength{\tabcolsep}{2.5pt}
\tabletypesize{\scriptsize}
\tablehead{
\colhead{$\mu$\textsuperscript{a}} & \colhead{$\sigma$} & \colhead{$\beta$} & \colhead{$A$\textsuperscript{b}} &
\colhead{$z^*_c$} & \colhead{$\ell_*$} & \colhead{$\gamma_1$} & \colhead{$\gamma_2$} &
\colhead{$\alpha$} & \colhead{$\delta$} & \colhead{$p2$} & \colhead{$-2\log L$} & \colhead{AIC}
}
\startdata
$2.030^{+0.007}_{-0.007}$ &
$0.203^{+0.005}_{-0.005}$ &
$0.006^{+0.005}_{-0.005}$ &
$2.5^{+13.5}_{-2.0} \times 10^4$ &
$1.75^{+0.26}_{-0.43}$ &
$0.14^{+0.59}_{-0.13}$ &
$0.53^{+0.11}_{-0.11}$ &
$1.11^{+1.26}_{-0.25}$ &
$0.15^{+0.03}_{-0.04}$ &
$3.2^{+1.8}_{-0.8}$ &
$-6.9^{+2.1}_{-1.8}$ &
$-8502$ & $-8482$
\enddata
\tablecomments{
\textsuperscript{a} Parameter values reported are the median of the best-fit values obtained for all 1000 Monte Carlo samples. Statistical uncertainties are reported in this table, computed as the 68\% containment region around the median value. \\
\textsuperscript{b} In units of $10^{-13}$\,Mpc$^{-3}$. \\
}
\end{deluxetable*}

\begin{figure*}
\centering

\gridline{
\fig{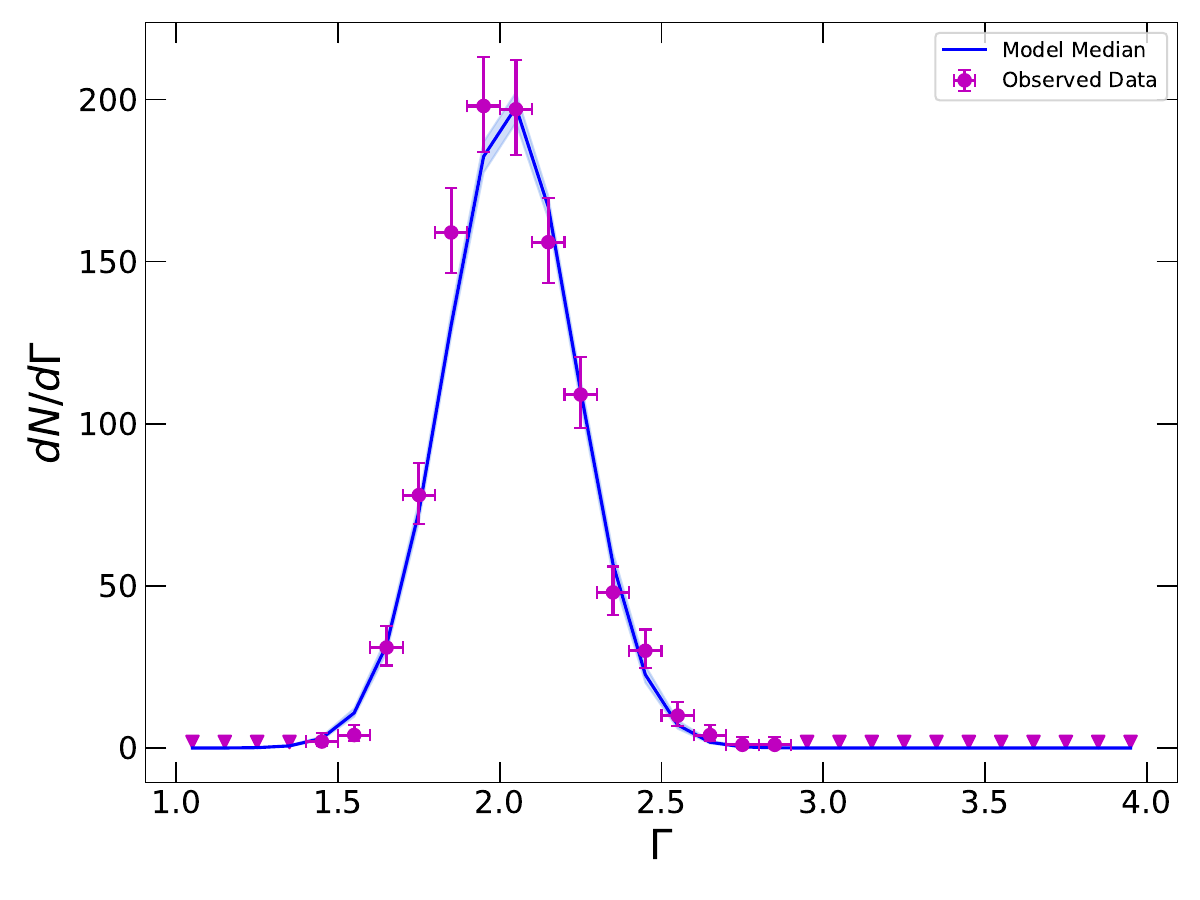}{0.48\textwidth}{(a)}
\fig{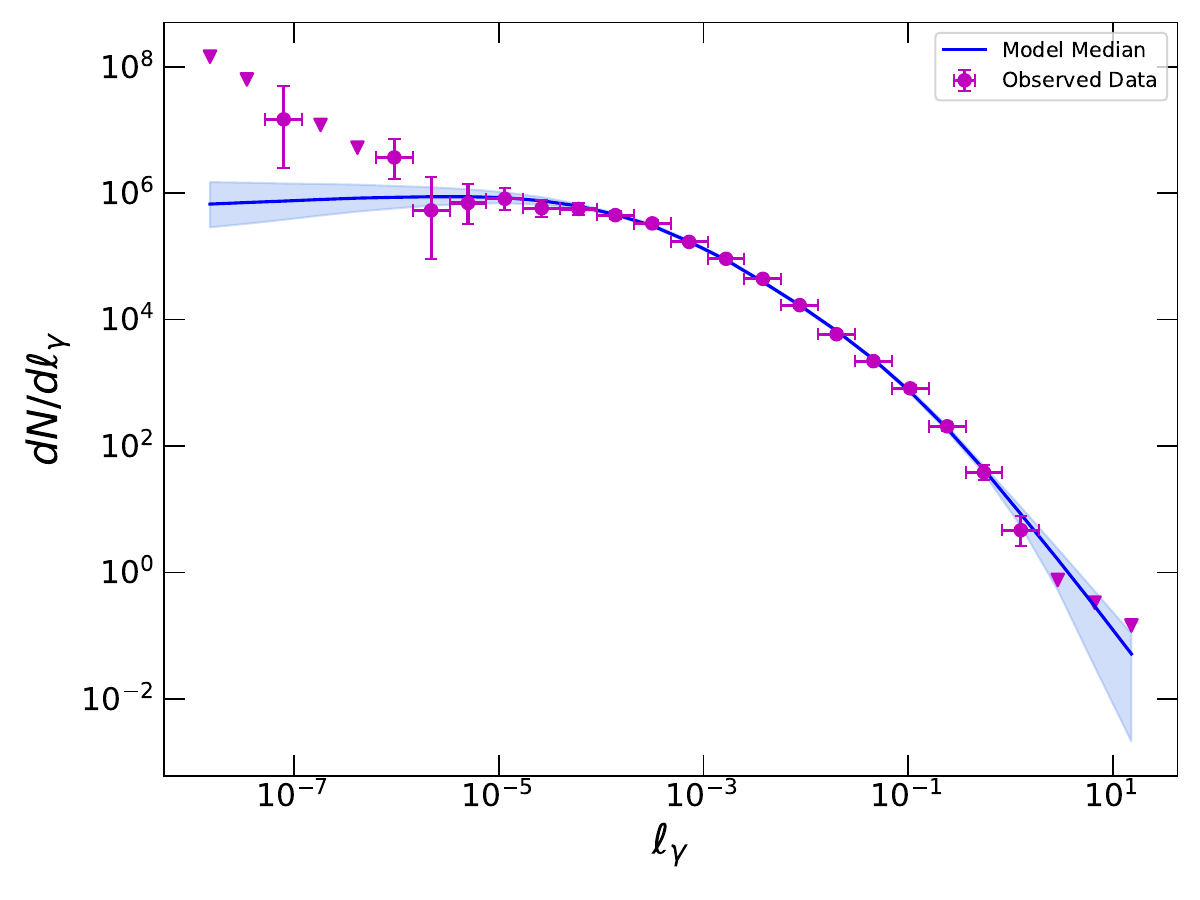}{0.48\textwidth}{(b)}
}

\gridline{
\fig{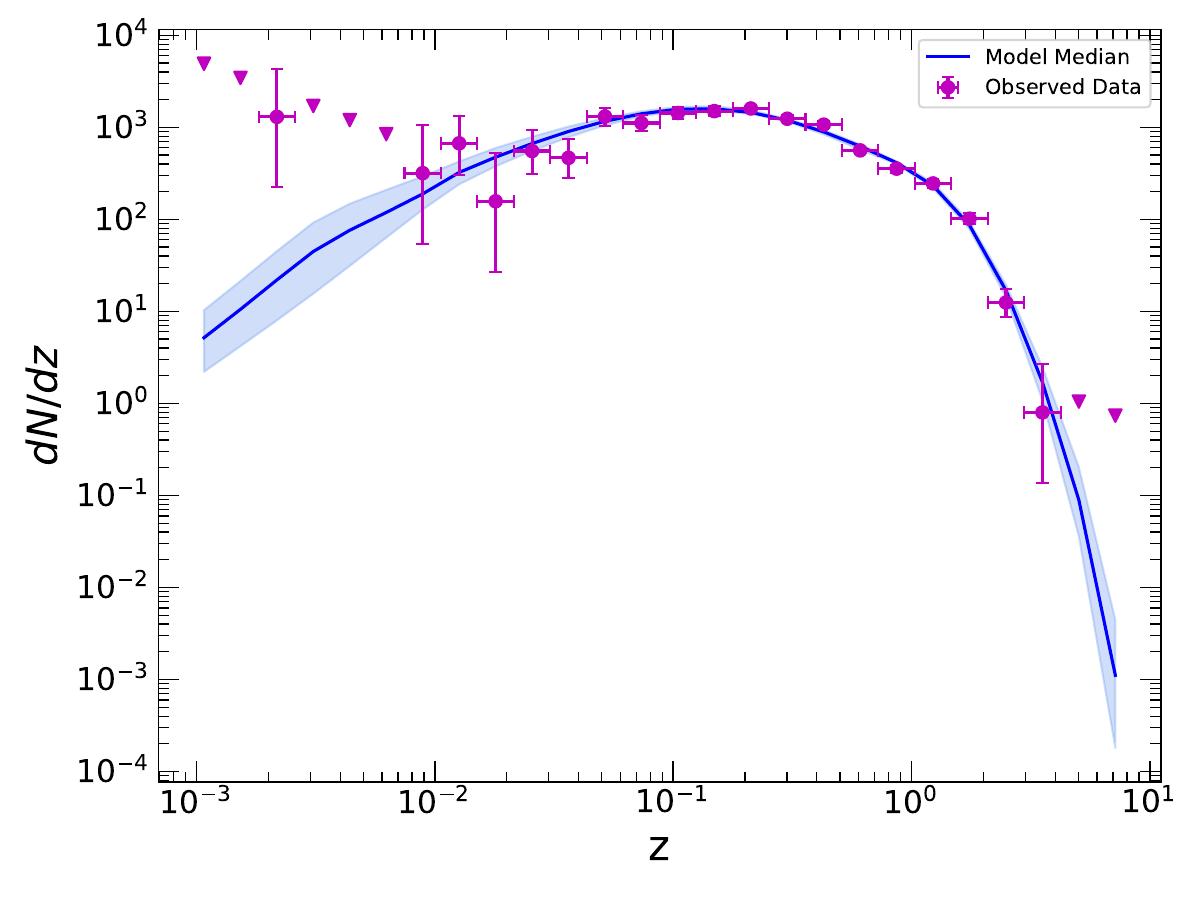}{0.48\textwidth}{(c)}
\fig{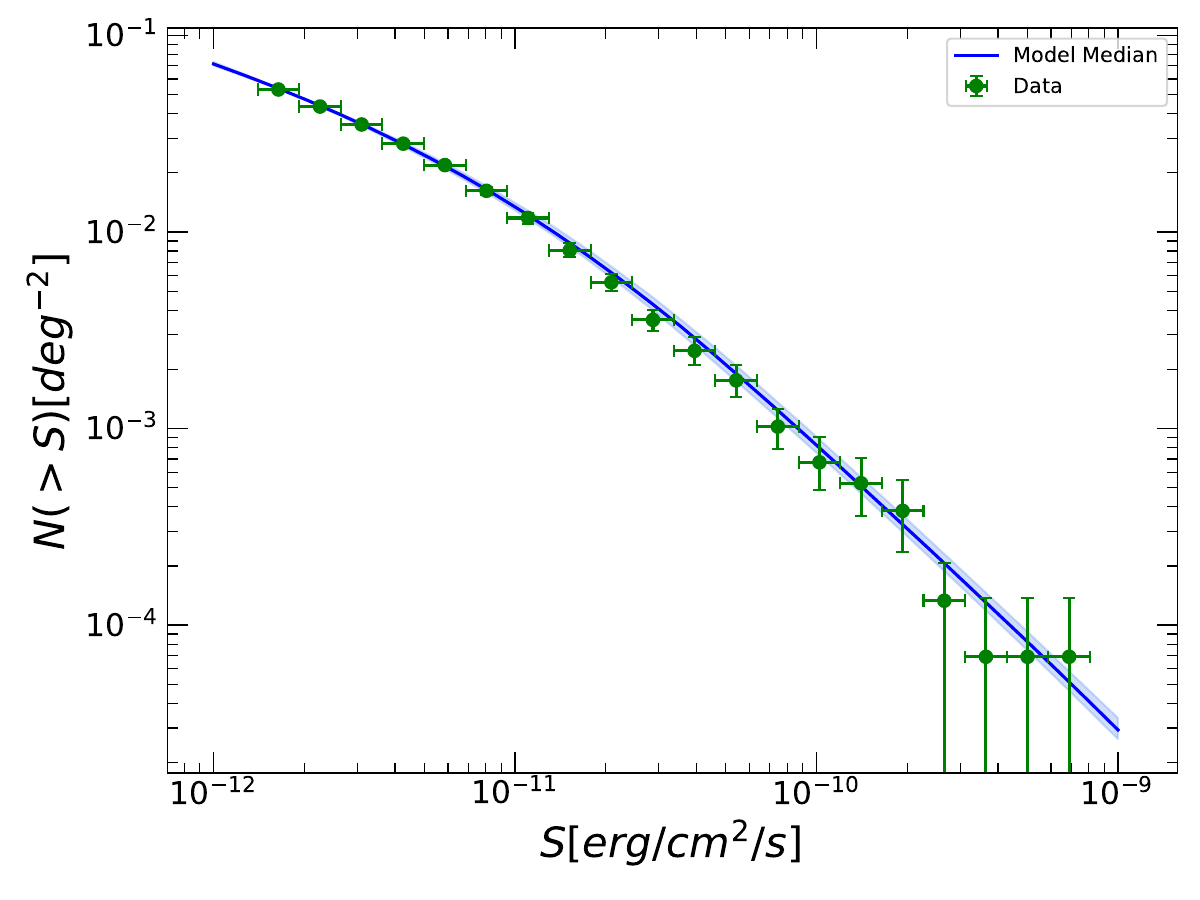}{0.48\textwidth}{(d)}
}

\caption{
(a) The observed photon-index distribution is given in the magenta points. The solid blue line, surrounded by the $1\sigma$ uncertainty region, denotes the median best-fit line calculated using the LF of 1000 Monte Carlo samples.
(b) The luminosity distribution is given in the magenta points (upper limits are shown in the downward triangles). The solid blue line, surrounded by the $1\sigma$ uncertainty region, denotes the median best-fit line.
(c) The observed redshift distribution is given in the magenta points (upper limits are shown in the downward triangles). The solid blue line, surrounded by the $1\sigma$ uncertainty region, denotes the median best-fit line.
(d) The intrinsic source-count distribution is given in the green points. The solid blue line, surrounded by the $1\sigma$ uncertainty region, denotes the median best-fit line.
}
\label{fig:ldde_panels}

\end{figure*}

\begin{table}
\begin{center}
\caption{Comparison of the LF Models.}
\begin{tabular}{cccc}
\hline
\hline
Model (i) & $-2\log L$ &  AIC$_i$ & $\Delta_i^{a}$ \\ 
\hline
\hline
PLE & -8453 & -8435   &  47  \\
PDE & -8457 & -8439   &  43   \\
LDDE$^{b}$ & -8502 & -8482 (best)  &  0   \\
\hline
\hline
\multicolumn{4}{l}{
\begin{minipage}{\linewidth} 
\tablenotetext{a}{$\Delta_i = AIC_i - AIC_{min(best)} $}
\tablenotetext{b}{The LDDE model has the minimum AIC.}
\end{minipage}
}
\end{tabular}
\label{Tab:BestModelInference}
\end{center}
\end{table}

The LDDE model describes a density-evolution scenario in which the redshift peak depends on the luminosity regime of the objects. This was found to be the best-fit model for our FSRQ sample \citep{Rajguru2025}. Since we already observed a luminosity dependence in the redshift peak of the previous distributions, we fit our sample with the LDDE model. Table \ref{tab:LDDE} shows the best-fit parameters of the LDDE model. Figure \ref{fig:ldde_panels} shows the agreement between the best-fit model lines and the observed photon-index, redshift distribution,  intrinsic source-count and luminosity distribution.

Comparing the $-2\ln L$ and AIC values in Tables \ref{Tab:BestModelInference}, we see that the LDDE model has the lowest values for both ($-2\ln L = -8502$ and AIC $=-8482$). The difference in AIC $\Delta_{\rm PLE-LDDE} = 47$ and $\Delta_{\rm PDE-LDDE} = 43$ gives strong evidence that the LDDE model is the best-fit model for BL Lacs. The relative likelihood (calculated as $p_{\rm i} = e^{-0.5\times(\rm AIC_{i} - AIC_{min})}$, where $\rm AIC_{min} = AIC_{LDDE}$) of the PLE and PDE models with respect to the LDDE model are $p_{\rm PLE}=6.2\times 10^{-11}$ and $p_{\rm PDE}=4.5\times 10^{-10}$ respectively. This means that the LDDE model is the best-fit model ($>6\sigma$). The large span in luminosities (3 orders of magnitude lower) and redshift ($z\sim4$) compared to previous studies has enabled us to establish the LDDE model as the best-fit model over the PLE and PDE models. Most of our best-fit parameters lie within the uncertainty ranges of the LDDE parameters given in \citet{Ajello2014}. Note that \citet{Ajello2014} used a smaller sample of 211 BL Lacs and had a different equation defining the LDDE model. We find that a simpler LDDE model can explain the BL Lac data.

\begin{figure}[htb!]
\centering
\includegraphics[width=\columnwidth]{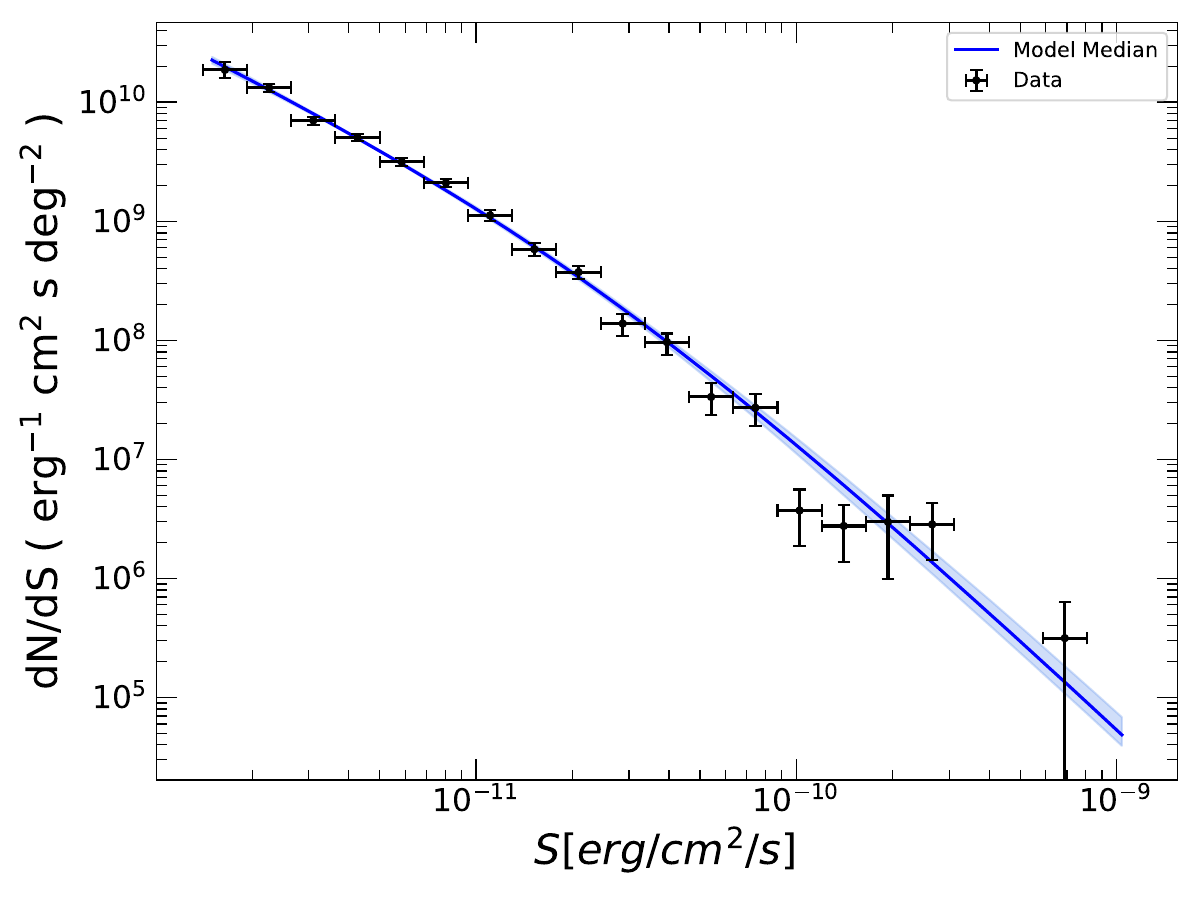}
\caption{The intrinsic differential source count distribution (\textit{logN-logS}) of BL Lacs plotted against the energy flux. The solid blue line represents the intrinsic \textit{logN-logS} as derived using the LDDE model, surrounded by the $\pm 1\sigma$ uncertainty region. The data points are calculated using Equation \ref{eqn:dNdS}. The error bars show the statistical uncertainty derived from 1000 Monte Carlo samples.}
\label{fig:dNdS}
\end{figure}

Given a particular flux interval, the intrinsic number of sources in that flux bin can be calculated using the equation:
\bbe
\frac{dN}{dS} = \frac{1}{\Omega(S_i)} \frac{N_i}{\Delta S_i}
\label{eqn:dNdS}
\ee
where $N_i=$ the number of BL Lacs in the $i$-th flux bin, $S_i =$ the $i$-th energy flux, $\Delta S_i = $ the flux interval centered at $S_i$, and $\Omega(S_i) =$ the sky coverage at flux $S_i$. Figure \ref{fig:dNdS} shows the intrinsic differential source-count distribution ($LogN-LogS$) of BL Lacs calculated from the best-fit LF in the solid blue line. We adopt equally spaced logarithmic bins in flux. For each logarithmically spaced bin, the differential source counts $dN/dS$ are calculated using the corresponding linear flux width, $\Delta S_i = S_{i,\rm max} - S_{i,\rm min}$.

\begin{figure*}[htb!]
\centering
\includegraphics[width=0.8\textwidth]{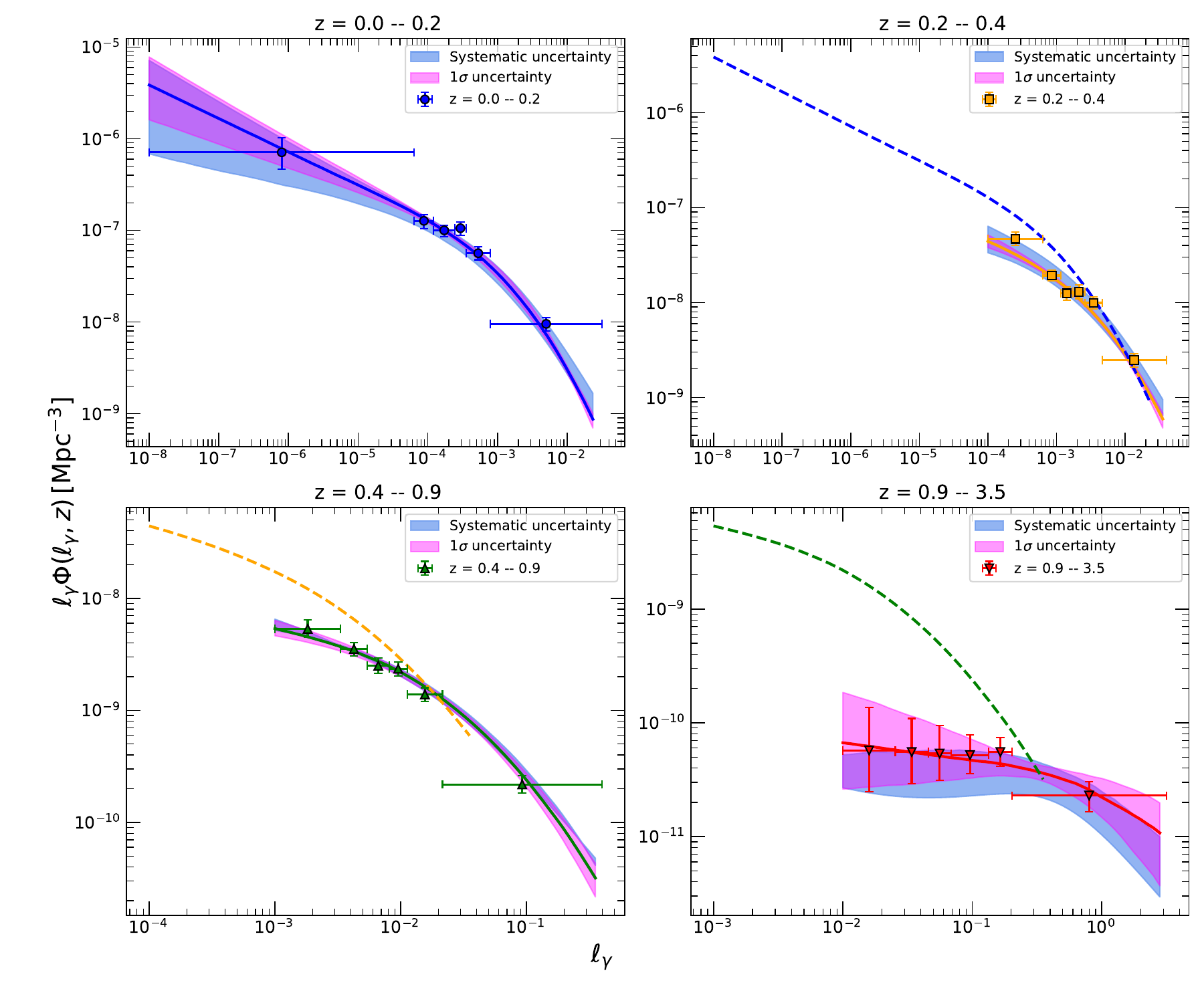}%
\caption{The LF of \textit{Fermi} BL Lacs subdivided into four bins in redshift, constructed using the $N_{obs}/N_{mdl}$ method. The solid lines depict the best-fit model (LDDE) in each redshift bin. The number density in the previous bin given is in the dashed lines for comparison. The statistical and systematic uncertainties are shaded in magenta and blue respectively. The systematic uncertainty band represents the envelope obtained by varying the assumptions regarding the redshift distribution of sources without measured redshifts and the sky coverage, whereas the statistical uncertainty quantifies the statistical variation arising from the stochastic assignment of redshifts within our $dN/dz$ prescription.}
\label{fig:LF_Evol_L}
\end{figure*}

\begin{figure*}[htb!]
\centering
\includegraphics[width=0.8\textwidth]{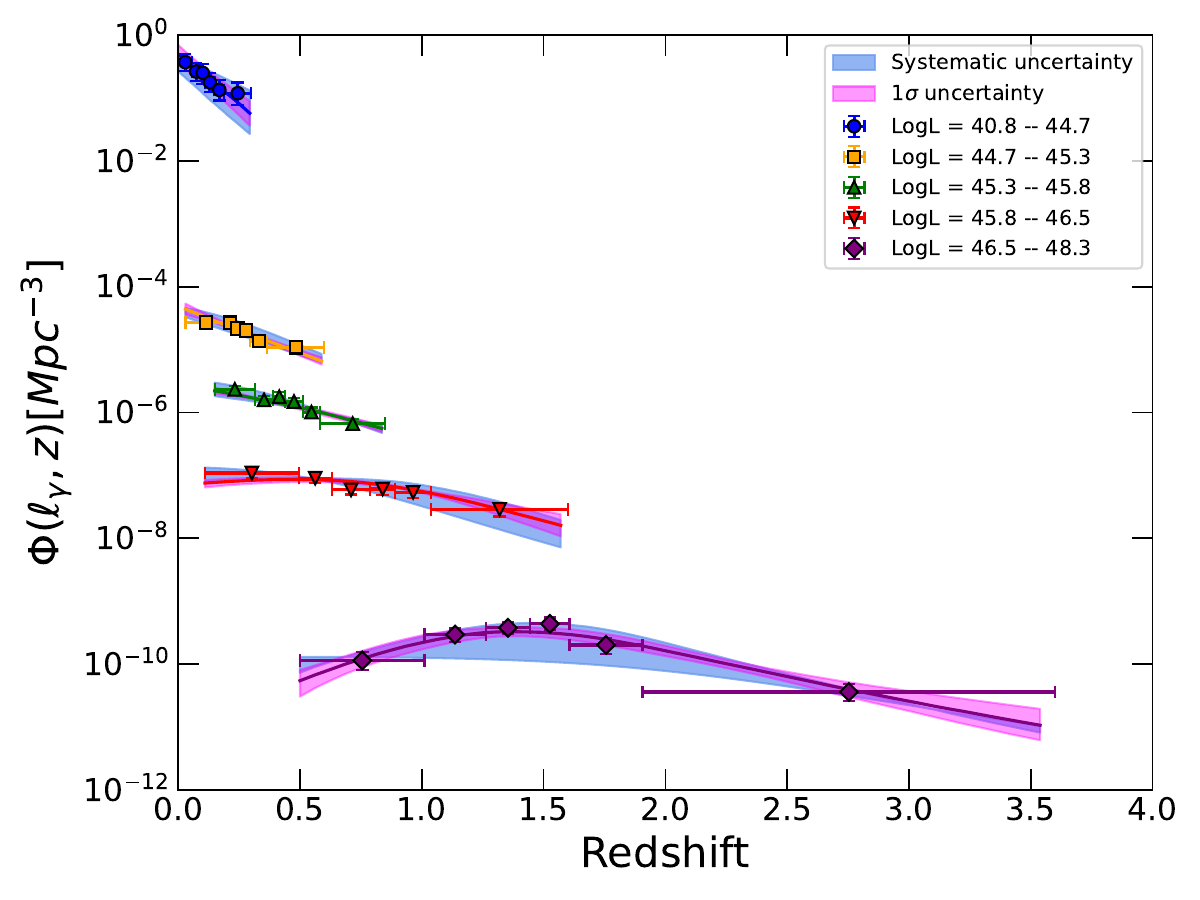}%
\caption{The evolution of \textit{Fermi} BL Lacs subdivided into five bins in luminosity, constructed using the $N_{obs}/N_{mdl}$ method by employing the best-fit (LDDE) model. The lowest luminosity bin shows negative evolution, whereas the highest luminosity bin shows positive evolution. Moreover, the peak in evolution is at higher redshift for a more luminous class; that is, this plot shows the luminosity dependence of evolution in density. The best-fit lines are given in different colors for different luminosity bins as described in the legend, and the data points are calculated using the $N_{obs}/N_{mdl}$ method. The statistical and systematic uncertainties are shaded in magenta and blue respectively.}
\label{fig:LF_Evol_z}
\end{figure*}

In Figure \ref{fig:LF_Evol_L} we show the LF as a function of luminosity in different redshift bins to demonstrate the evolution of the LF in different redshift intervals. We divide our sample into 4 redshift bins, keeping the number of sources in each bin roughly the same. Figure \ref{fig:LF_Evol_L} shows that both the shape and normalization of the BL Lac LF vary across the different redshift intervals, demonstrating the redshift-dependent evolution of the population. In particular, the relative contribution of sources at different luminosities changes with increasing redshift, as described by the best-fit LDDE model. In the lowest redshift bin ($z = 0.0 - 0.2$), we see that the contribution of low-luminosity objects is the highest, as indicated by the blue solid line. As we look into the next two redshift intervals, i.e. $z = 0.2 - 0.4$ and $z = 0.4 - 0.9$, we see that the slope of the LF flattens, indicating an increase in contribution from comparatively higher luminosity objects. The major episode of evolution for BL Lacs occurs at $z>0.9$ (the last redshift bin), where the shape of the LF changes such that the contribution of the high-luminosity sources matches the contribution of low-luminosity sources.

Figure \ref{fig:LF_Evol_z} shows the luminosity dependence of the LF evolution. The sample is divided into 5 luminosity intervals by keeping the number of objects in each interval roughly the same. We see a change in the trend of the slopes of LF with luminosity regime. The lowest luminosity bin (blue line) shows a negative evolution, where number of sources were lower in the past. On the other hand, highest luminosity regime shows a positive evolution, where the number of sources were higher in the past compared to the present. We can see this behavior in the most luminous class (purple solid line), where the number densities initially increase till a certain redshift and then starts to decrease again. The redshift at which the evolution changes from positive to negative is the turnover redshift ($z_c^*$). The turnover redshift of the most luminous class of our sample ($L_\gamma = 10^{48}$ erg/s) is $z_c^* = 1.75^{+0.26}_{-0.43}$, which is in agreement (within uncertainties) with that found in \citet{Ajello2014} $z_c^* = 1.34$. The parameter governing the evolution of the redshift peak is $\alpha = 0.15^{+0.03}_{-0.04}$. Thus, the evolution changes from a negative to a positive evolution as we move towards higher luminosities.

The results shown in this section use the EBL-corrected luminosities defined in Section \ref{sec:EBL}; the corresponding analysis without the EBL correction is presented in Appendix \ref{sec:NoEBL_LF}.

\subsection{The Local LF}
\label{localLF}

\begin{figure*}[htb!]
\centering
\includegraphics[width=0.8\textwidth]{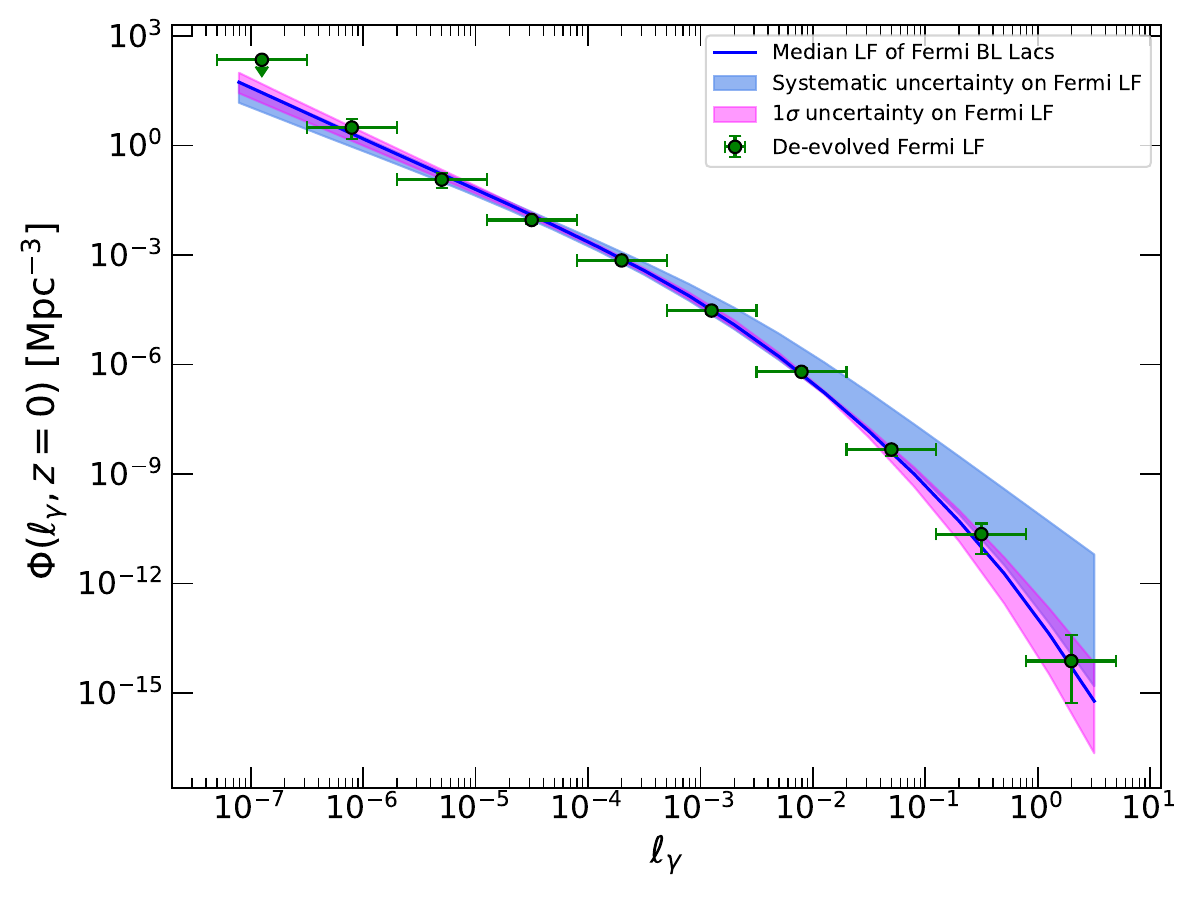}
\caption{The local LF of \textit{Fermi} BL Lacs, de-evolved at $z=0$, is denoted by the green points, calculated using the $1/V_{\rm max}$ method for the 1000 Monte Carlo samples. For each data point, the error bars are calculated as the 68\% confidence interval around the median. The magenta hatched area is the $\pm1\sigma$ uncertainty of the LF. The systematic uncertainty is shaded in blue. The blue solid line depicts the best-fit local LF, given by the median of 1000 samples.}
\label{fig:LF_DeEvol}
\end{figure*}

The local LF is estimated by de-evolving the number densities to a redshift $z=0$ according to the best-fit model. We take into account the luminosity-dependence while de-evolving the number densities since our best-fit model is the LDDE model. Using the modified version of the $1/V_{\rm max}$ method \citep{DellaCeca2008}, which was originally given by \citet{Schmidt1968}, we calculate the de-evolved LF ($\Phi(\ell_\gamma,z=0)$)  of BL Lacs as:
\bbe\label{eq:vmax}
V_{\rm max, i} = \int^{z_{max} = \, 10}_{z_{min} = \, 0} \Omega(\ell_{\gamma,i},z, \Gamma_i) \frac{e'(z,\ell_{\gamma,i})}{e'(z_{min},\ell_{\gamma,i})}
  \frac{dV_c}{dz}dz 
\ee
\bbe\label{eq:dev}
\Phi(\ell_\gamma, z=0) = \frac{1}{\Delta \ell_\gamma} \sum_{i\in\mathrm{bin}}\frac{1}{V_{\rm max, i}}
\ee
where $\ell_{\gamma,i}$ is the scaled luminosity of source $i$, $\Omega(\ell_{\gamma,i},z,\Gamma_i)$ the sky coverage
and $e'(z,\ell_{\gamma,i})$ the evolution term of Eq.~\ref{eq:10} normalized
(through $e'(z_{min},\ell_{\gamma,i})$) at the redshift $z_{min} = 0$ to which
the LF is to be de-evolved. To estimate the effective accessible volume to which a source can be detected ($V_{\rm max, i}$) by the survey, we integrate Eq.~\ref{eq:vmax} to a maximum redshift $z_{max} = 10$, and compute at every redshift the $\Omega(\ell_{\gamma,i},z,\Gamma_i)$ given the source photon index and luminosity. We use the sample of 1035 BL Lacs, without imposing an additional redshift cut to derive the local LF. As described in Section \ref{sec:Redshifts}, we generate 1000 bootstrap realizations by randomly drawing 1035 sources with replacement from the parent sample. Consequently, each realization contains 1035 source entries, but not necessarily 1035 distinct sources, as some sources may be selected multiple times while others may not be selected. For sources with measured redshifts, the measured redshift is retained in every Monte Carlo realization. For sources without measured redshifts, we use the redshift assignment following the procedure described in Section \ref{sec:Redshifts}. Thus, each Monte Carlo realization contains a sample of 1035 sources, with a redshift assigned to every source. Finally, we bin the sources according to their luminosity. We define equally spaced logarithmic bins in scaled luminosity $\ell_\gamma$ and estimate the local LF by summing the $1/V_{\rm max, i}$ factor of the sources in each luminosity bin (Eq.~\ref{eq:dev}).

In Figure \ref{fig:LF_DeEvol} we see that the local LF is described by a double power-law shape. The knee of this distribution occurs around a luminosity $L_\gamma \sim 10^{45} \rm erg/s$. We see that the best-fit line derived from the LDDE model agrees with the de-evolved data points calculated using the $1/V_{\rm max}$ method.

\section{Discussions} \label{sec:conc}

\begin{figure}[htb!]
\centering
\includegraphics[width=\columnwidth]{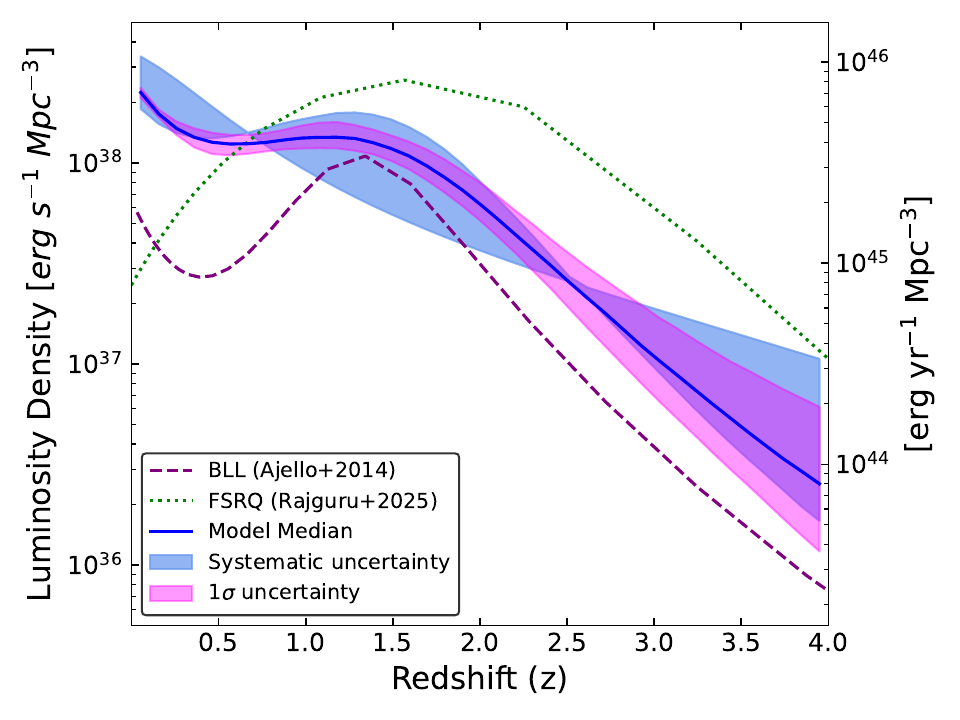}
\caption{The luminosity density of \textit{Fermi} BL Lacs. The best-fit line of our model is represented by the blue line (median of 1000 Monte Carlo samples), surrounded by the $\pm1\sigma$ uncertainty region shaded in magenta and the systematic uncertainty shaded in blue. The green dotted line shows the luminosity density of FSRQs obtained using best-fit parameters from \citet{Rajguru2025}. The purple dashed line depicts the luminosity density of BL Lacs obtained from \citet{Ajello2014}.}
\label{fig:LumDens}
\end{figure}

\begin{figure}[htb!]
\centering
\includegraphics[width=\columnwidth]{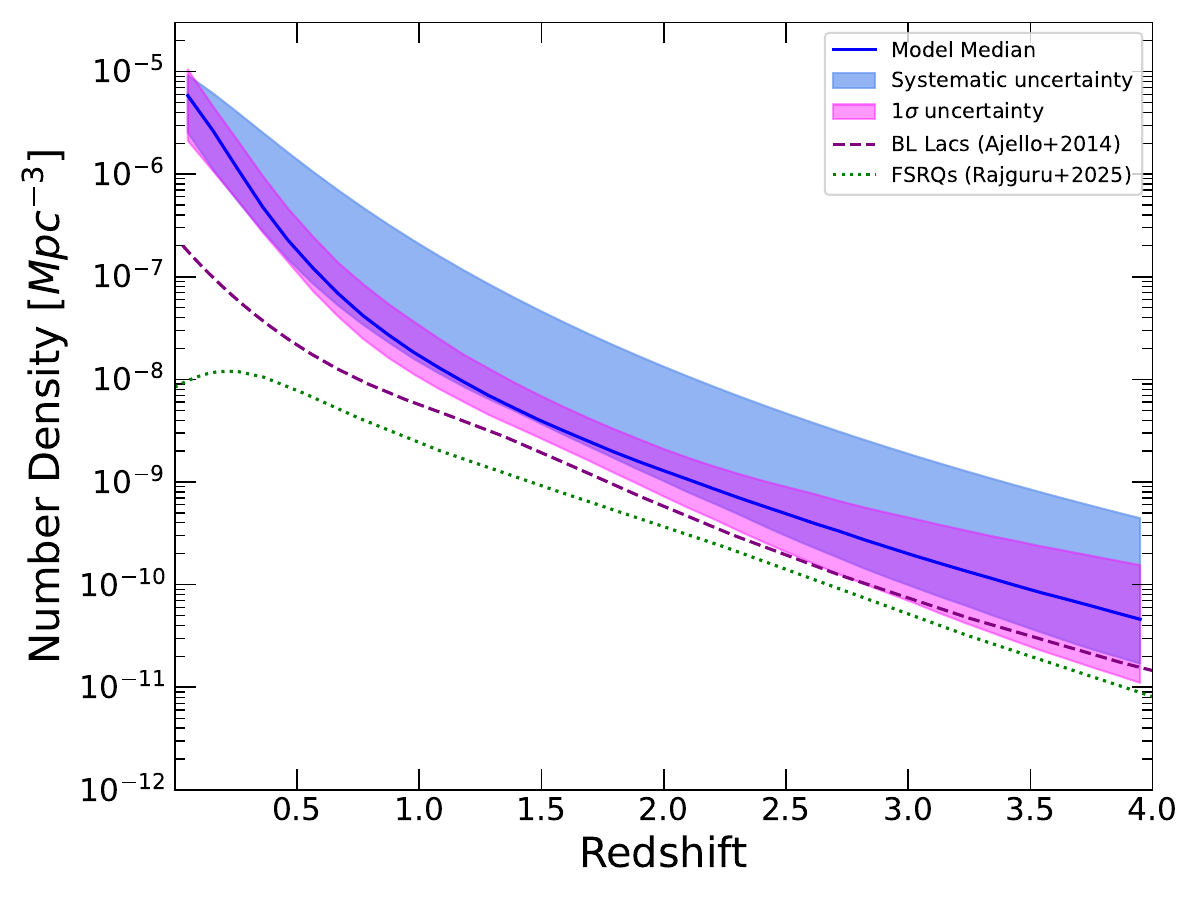}
\caption{The number density of \textit{Fermi} BL Lacs. The best-fit line of our model is represented by the blue line (median of 1000 Monte Carlo samples), surrounded by the $\pm 1\sigma$ uncertainty region shaded in magenta and the systematic uncertainty shaded in blue. The green dotted line shows the number density of FSRQs obtained using best-fit parameters from \citet{Rajguru2025}. The purple dashed line depicts the number density of BL Lacs obtained from \citet{Ajello2014}.}
\label{fig:NumDens}
\end{figure}

In this study, we utilized the largest sample of BL Lacs to generate the gamma-ray LF. The present sample of 1035 BL Lacs is nearly five times larger than the 211 object sample used by \citet{Ajello2014}. The sample consists of 1035 BL Lacs detected by \textit{Fermi}-LAT at high Galactic latitudes $|b|>20^\circ$ and in the energy range of 100 MeV-1 TeV. We fit analytical models to our data using the Maximum Likelihood approach for optimization, and find the best-fit model. In this section, we discuss the main implications of our work.

The LDDE model fits the data better than the PDE or PLE models. In the literature, an open question is the evolution of BL Lacs, as different studies report different evolutionary trends \citep[positive, negative or no evolution, e.g., ][]{2013Marcha, Bhattacharya2009_GLF, Beckmann2003}. We find that the evolution of BL Lacs depends on the luminosity range to which they belong. Low-luminosity BL Lacs ($L_\gamma < 10^{46} \rm erg/s$) evolve negatively, whereas high-luminosity BL Lacs evolve positively. This luminosity-dependence of the density evolution is also seen in the case of FSRQs \citep{Rajguru2025}. We note that mathematically the best-fit LDDE model in our study is simpler than the best-fit LDDE model optimized in \citet{Ajello2014} for \textit{Fermi} BL Lacs, and more similar to the LDDE model describing the evolution of FSRQs \citep{Ajello2012, Rajguru2025}, indicating a connection between the two blazar subclasses.

Figure \ref{fig:LF_DeEvol} shows the local LF (at $z=0$) as described by a double power law. \citet{Ajello2014} had also found a similar nature for 211 Fermi BL Lacs. \citet{2013Marcha} used a sample of 130 BL Lacs and discovered a break in the radio LF. The flattening of the slope at the low-luminosity end is seen in FSRQs as well \citep[e.g., ][]{Ajello2012, Rajguru2024}, and can be reconciled with the fact that BL Lacs are beamed sources. Beaming has been established to flatten the slope at lower luminosities as a result of the spread of bulk Lorentz factors that beam lower-luminosity objects to higher apparent luminosities \citep{UrryShafer1984,Urry1991,Ajello2012}.

We know that the luminosity density ($\rho_L(z)$) and the number density ($n(z)$) can be derived once we know the LF of a source class, as,
\bbe
\rho_L(z) = \int_{\ell_{\gamma,\min}}^{\ell_{\gamma,\max}} L_\gamma\,\Phi(\ell_\gamma,z)\,d\ell_\gamma \\
n(z) = \int_{\ell_{\gamma,\min}}^{\ell_{\gamma,\max}} \Phi(\ell_\gamma,z)\,d\ell_\gamma 
\ee
where the integration limits are $\ell_{\gamma,min}$ = $5.71\times10^{40}$\,erg s$^{-1} /L_0$ and $\ell_{\gamma,max}$ = $1.76\times10^{48}$\,erg s$^{-1} /L_0$. 

Figure \ref{fig:LumDens} shows the luminosity density of BL Lacs obtained from the best-fit LF of our study. We find that the BL Lacs luminosity density peaks twice, as was found in \citet{Ajello2014} depicted in the purple dashed line. One of the peaks is around $z\sim 1-2$, which aligns with our findings for FSRQs shown in the green dotted line \citep{Rajguru2025}, and is at an epoch when the star formation rate also peaked \citep[e.g.,][]{HopkinsBeacom2006,MadauDickinson2014,DaCunha2023}. This underscores the link between the host galaxy and the active nucleus and holds true for FSRQs and BL Lacs. Another interesting fact is that the luminosity density of BL Lacs continues to increase toward the present epoch ($z=0$). This is mainly due to the low-luminosity sources that are injecting energy into the universe at the current time. 

\begin{figure}[htb!]
\centering
\includegraphics[width=\columnwidth]{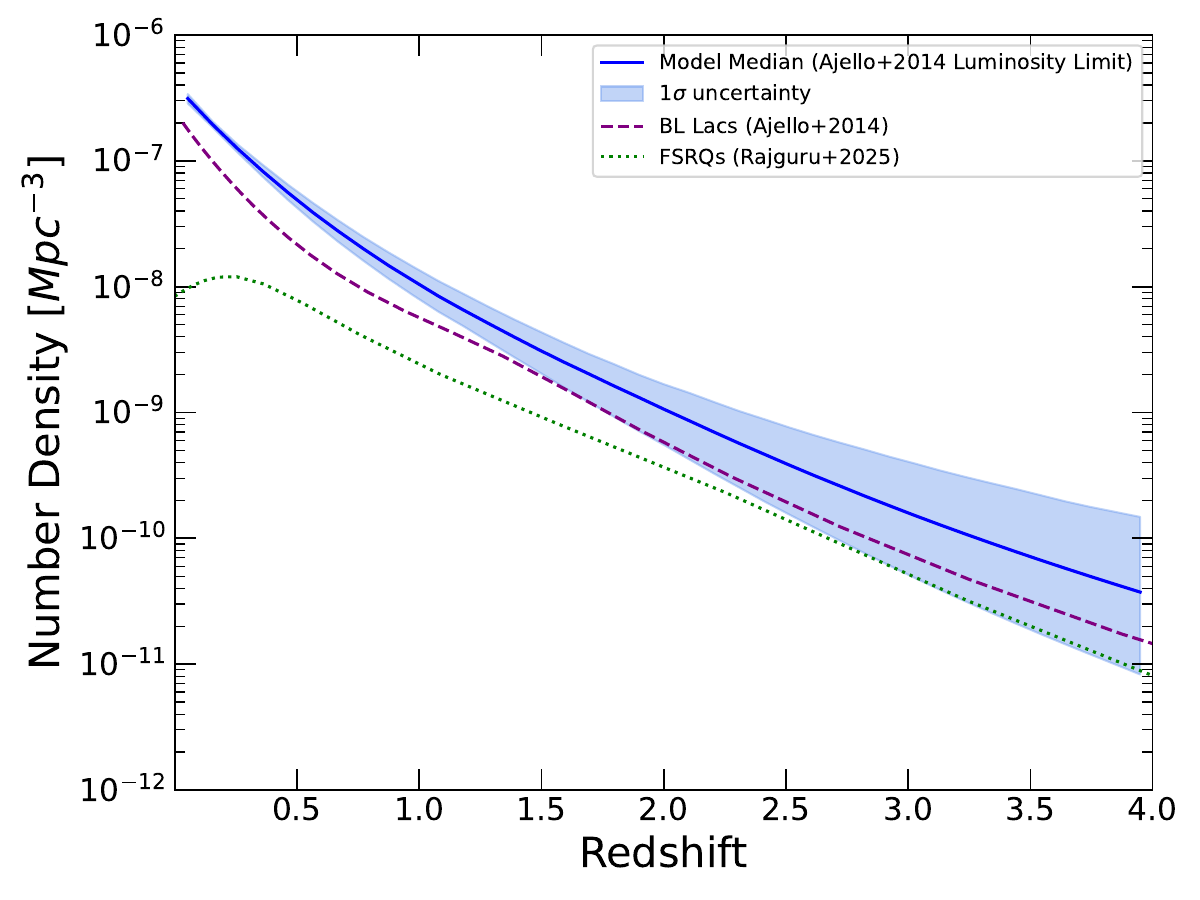}
\caption{The number density of \textit{Fermi} BL Lacs using the luminosity limits of \citet{Ajello2014} is shown for comparison. The best-fit line of our model is represented by the blue line (median of 1000 Monte Carlo samples), surrounded by the $\pm 1\sigma$ uncertainty region shaded in magenta. The purple dashed line depicts the number density of BL Lacs obtained from \citet{Ajello2014}. The number densities in both the works overlap within uncertainties for major part of the evolutionary history. The green dotted line shows the number density of FSRQs obtained using best-fit parameters from \citet{Rajguru2025}. We see the similarity of the densities of BL Lacs and FSRQs.}
\label{fig:NumDens_A14}
\end{figure}

The number density of BL Lacs, calculated using our best-fit LF model, is displayed in Figure \ref{fig:NumDens} in the solid blue line. We see that this agrees well with the number density of BL Lacs reported by \citet{Ajello2014}, with a higher number density at lower redshifts. This also aligns with the FSRQ number density reported by \citet{Rajguru2025}. In \citet{Rajguru2025}, it was noted that the number densities of FSRQs and BL Lacs suggested similar evolution. We compare in Figure \ref{fig:NumDens_A14} the number densities of BL Lacs from our study to those given in \citet{Ajello2014} by integrating both best-fit models within the luminosity limits given in that paper ($L_{\gamma, \rm min} = 7\times 10^{43} \rm erg/s$ and $L_{\gamma, \rm \, min} = 10^{52} \rm \, erg/s$). The two densities overlap almost entirely, indicating that the increase in number density in current epochs is due to the large number of low-luminosity sources in our sample. 

In this study, we confirm that the evolution of BL Lacs and FSRQs is indeed similar. The similarity in the number densities of BL Lacs and FSRQS can be due to two prominent reasons \citep{Rajguru2025}. 
First, the BL Lac sample may contain `masquerading' BL Lacs or `blue FSRQs', which are intrinsically FSRQs having their emission lines swamped out by an overpowering contribution from the jet at optical wavelengths \citep{Ghisellini2012_blueFSRQ,Padovani2012,Giommi2013, padovani2019, Rajagopal2020, Rajguru2024}. Since the classification of BL Lacs and FSRQs is done mainly on the basis of the width of emission lines, such objects are categorized as BL Lacs \citep{Ruan2014, Mishra2021}, and can exhibit an evolutionary trend similar to FSRQs.

Second, the similarity in evolution of the two subclasses of blazars can be explained if there is a strong genetic link between BL Lacs and FSRQs, both being jetted accreting black holes, with differences in their Doppler beaming, accretion rate or available circumnuclear matter \citep{Vagnetti1991_LinkBLL, cavaliere2002,boettcher2002, FermiBlazarDivide2009}. BL Lacs may represent the later stages of FSRQs when there is a paucity of cold gas near the center of galaxies, and fewer galaxy mergers in the universe to facilitate quasar activity \citep{Sanders1988, DiMatteo2005, Hopkins2008, KulkarniLoeb2012}. Thus, BL Lacs and FSRQs may show similar luminosity-dependent density evolution despite their differences in observed properties. 

\section{Summary}    \label{sec:summary}

This work uses 1035 BL Lacs, observed by \textit{Fermi}-LAT in the 100 MeV - 1 TeV energy range, to study the gamma-ray LF and the evolution of BL Lacs. The main findings are as follows:
\begin{enumerate}
    \item We utilize the largest sample of BL Lacs (1035) available in the literature to construct their LF. The sample includes low-luminosity objects down to $L_\gamma = 5.71 \times 10^{40} \rm \, erg/s$ and redshifts up to $z\sim 4$. This enables us to study the evolutionary behavior of low-luminosity objects up to high redshifts.
    \item We find that BL Lacs follow a luminosity-dependent density evolution, where we observe a change in the evolution depending on the luminosity of the object (Figure \ref{fig:LF_Evol_z}). We see negative evolution for low-luminosity objects ($L_\gamma<10^{46}\rm erg/s$) and positive evolution for high-luminosity objects. This means that low-luminosity BL Lacs are common at present times but high-luminosity objects were common in the past and are declining in numbers as we look into current time. This is also shown in Figure \ref{fig:LF_Evol_L}.
    \item The local LF of BL Lacs follows a double power-law shape, which is expected for beamed sources \citep{UrryShafer1984}.
    \item The luminosity density of BL Lacs shows peaks at redshifts $z\sim1.5$ and $z=0$. The number density of BL Lacs is currently peaking at $z=0$. Moreover, we confirm that BL Lacs and FSRQs seem to show similar evolution in their number densities. This can be due to contamination of the BL Lac sample by `masquerading' BL Lacs having overwhelming jet contributions in the optical regime, or an intrinsic similarity in the evolution of jetted accreting black holes.
\end{enumerate}

\begin{acknowledgments}
We thank the referee for comments that helped in improving the manuscript. G.R. and M.A. acknowledge the NASA contract 80NSSC24K0284 for this project. This research has used redshifts available in the online data base ZBLLAC (\url{https://web.oapd.inaf.it/zbllac/index.html}).

LM acknowledges that this work was supported by the the Initiative and Networking Fund of the Helmholtz Association under the Helmholtz Investigator Groups Programme, call 2025 (VH-NG-21-01). LM acknowledges support from DESY (Zeuthen, Germany), a member of the Helmholtz Association HGF.

\end{acknowledgments}

\appendix
\restartappendixnumbering

\section{The Sample of BL Lacertae objects}
\label{sec:appx_sample}

Table \ref{tab:BLL_List} provides the composition of the $\gamma$-ray sample of BL Lacs used in our study.

\begin{deluxetable*}{llllllllllll}[h!]
\tablecaption{Composition of the Gamma-Ray Sample of BL Lacertae objects$^{a}$.}
\label{tab:BLL_List}
\tablewidth{0pt}
\tabletypesize{\scriptsize}
\tablehead{
\colhead{Source Name} & \colhead{4FGL Name} & \colhead{RA$^{b}$ (J2000)} & \colhead{DEC$^{b}$ (J2000)} & \colhead{GLAT$^{b}$} & \colhead{eflux$^{c}$} & \colhead{eflux Unc.$^{c}$} & \colhead{CLASS} &
\colhead{Redshift} & \colhead{PL Index} & \colhead{PL Index Unc.} & \colhead{Counterpart Name}
}
\startdata
PS J1353.3$+$5602 & 4FGL J1353.4$+$5600 & 208.33 & 56.05 & 59.0152 & 1.29$\times 10^{-12}$ & 5.65$\times 10^{-13}$ & BLL & 0.404 & 2.085 & 0.152 & RX J1353.4$+$5601 \\
PS J1202.3$+$4443 & 4FGL J1202.4$+$4442 & 180.58 & 44.72 & 69.8447 & 1.29$\times 10^{-12}$ & 3.28$\times 10^{-13}$ & BLL & 0.297 & 2.544 & 0.160 & B3 1159$+$450 \\
PS J1223.8$+$4650 & 4FGL J1223.8$+$4649 & 185.96 & 46.85 & 69.5468 & 1.36$\times 10^{-12}$ & 4.87$\times 10^{-13}$ & BLL & 0.261 & 2.216 & 0.208 & RX J1223.8$+$4651 \\
PS J0944.1$+$5558 & 4FGL J0944.2$+$5557 & 146.05 & 55.97 & 45.9526 & 1.37$\times 10^{-12}$ & 3.27$\times 10^{-13}$ & BLL & --- & 2.219 & 0.142 & NVSS J094441$+$555752 \\
PS J0946.1$+$4733 & 4FGL J0946.0$+$4735 & 146.54 & 47.56 & 48.7039 & 1.43$\times 10^{-12}$ & 4.52$\times 10^{-13}$ & BLL & 0.569 & 2.087 & 0.217 & RX J0946.0$+$4735 \\
\enddata
\tablecomments{
\textsuperscript{a} Table \ref{tab:BLL_List} is published in its entirety in the machine-readable format. A portion is shown here for guidance regarding its form and content. \\
\textsuperscript{b} In units of deg. \\
\textsuperscript{c} Energy flux over 100 MeV - 1 TeV (in units of $\rm erg\, cm^{-2}\, s^{-1}$). \\
}
\end{deluxetable*}

\section{$L_\gamma-$\lowercase{$z$} Region Spanned by the Data}
\label{sec:LzPlot}

\begin{figure}[htb!]
\centering
\includegraphics[width=\columnwidth]{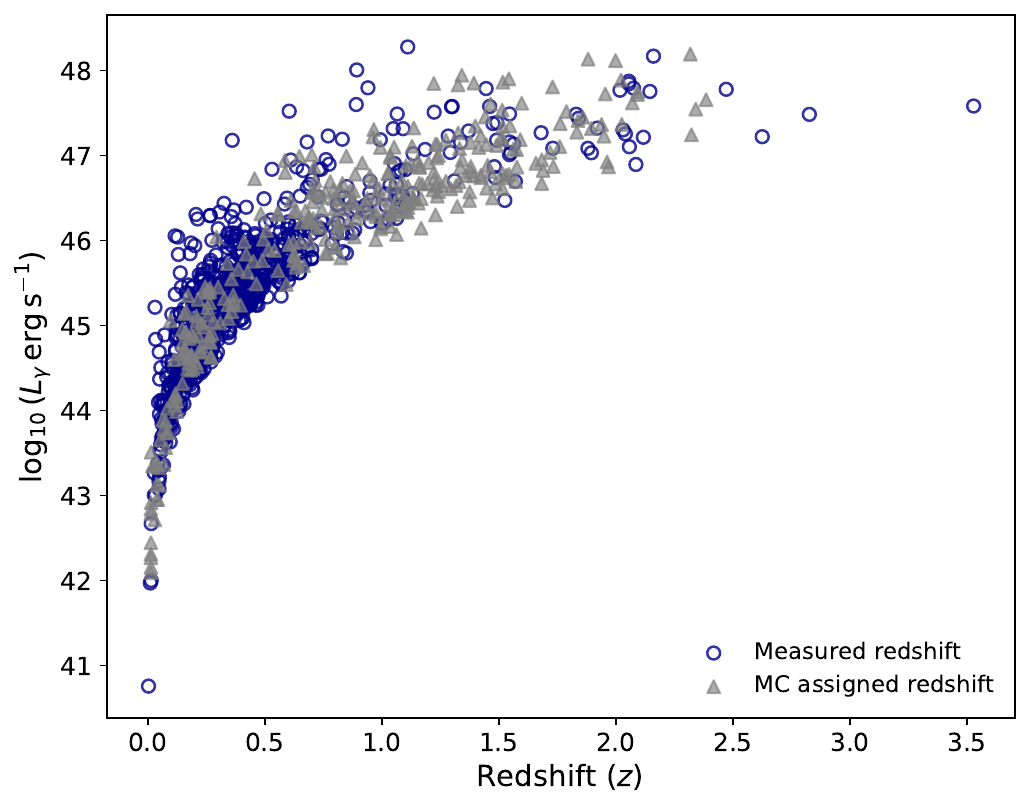}
\caption{The luminosity $L_\gamma$ vs redshift plot of the BL Lacs in one representative sample among the 1000 samples. The blue open circles represent the sources having a measured redshift, whereas the gray triangles are the sources having a redshift value assigned through the MC approach (see Section \ref{sec:Redshifts}). The plot also illustrates the luminosity–redshift parameter space spanned by the sample.}
\label{fig:LzPlot}
\end{figure}

To illustrate the region of $L_\gamma-z$ covered by the BL Lac data, we have plotted the $L_\gamma$ vs $z$ for one representative sample in Figure \ref{fig:LzPlot}. We distinguish sources with measured redshifts from those without measured redshifts using different colors and symbols. For sources lacking a measured redshift, a redshift value is assigned through the Monte Carlo (MC) procedure described in Section \ref{sec:Redshifts}.

\section{Comparison of LF With and Without EBL Correction}
\label{sec:NoEBL_LF}

For comparison, we also repeated the LF analysis without correcting the luminosities for EBL absorption for our best-fit LDDE model. Mathematically, this means using $C_{\rm EBL} = 1.0$ in Eqn. \ref{eqn:luminosity} to calculate luminosities. The resulting LF parameters are shown in Table \ref{tab:LDDE_noEBL}. Comparing Table \ref{tab:LDDE} with Table \ref{tab:LDDE_noEBL} we see that the parameters are consistent with those obtained by using the EBL-corrected luminosities within their uncertainties. For example, $z^*_c$ changes from $1.75^{+0.26}_{-0.43}$ to $1.87^{+0.26}_{-0.47}$, and $\alpha$ changes from $0.15^{+0.03}_{-0.04}$ to $0.16^{+0.03}_{-0.05}$. The overall evolutionary behavior of the LF remain unchanged within the uncertainties. We therefore conclude that including the EBL correction does not appreciably affect the main conclusions of this work.

\begin{deluxetable*}{lllllllllllll}
\tablecaption{Best-fit parameters of the LDDE model fit on luminosities not including EBL-correction.}
\label{tab:LDDE_noEBL}
\tablewidth{\textwidth}
\setlength{\tabcolsep}{3pt}
\tabletypesize{\scriptsize}
\tablehead{
\colhead{$\mu$\textsuperscript{a}} & \colhead{$\sigma$} & \colhead{$\beta$} & \colhead{$A$\textsuperscript{b}} &
\colhead{$z^*_c$} & \colhead{$\ell_*$} & \colhead{$\gamma_1$} & \colhead{$\gamma_2$} &
\colhead{$\alpha$} & \colhead{$\delta$} & \colhead{$p2$} & \colhead{$-2\log L$} & \colhead{AIC}
}
\startdata
$2.046^{+0.009}_{-0.009}$ & $0.198^{+0.004}_{-0.004}$ & $0.012^{+0.005}_{-0.005}$  & $2.5^{+5.0}_{-1.6} \times 10^4$ & $1.87^{+0.26}_{-0.47}$ & $0.28^{+0.46}_{-0.21}$  & $0.56^{+0.07}_{-0.08}$  & $1.02^{1.06}_{0.02}$ &
$0.16^{+0.03}_{-0.05}$ & $3.3^{+1.7}_{-0.8}$ & $-7.7^{+2.2}_{-1.7}$  & -8690 & -8670
\enddata
\tablecomments{
\textsuperscript{a} Parameter values reported are the median of the best-fit values obtained for all 1000 Monte Carlo samples. Statistical uncertainties are reported in this table, computed as the 68\% containment region around the median value. \\
\textsuperscript{b} In units of $10^{-13}$\,Mpc$^{-3}$. \\
}
\end{deluxetable*}

\section{Markov chain Monte Carlo (MCMC) Analysis using the LDDE Model}
\label{sec:LDDE_MCMC}
\begin{figure*}[htb!]
\centering
\includegraphics[width=\textwidth]{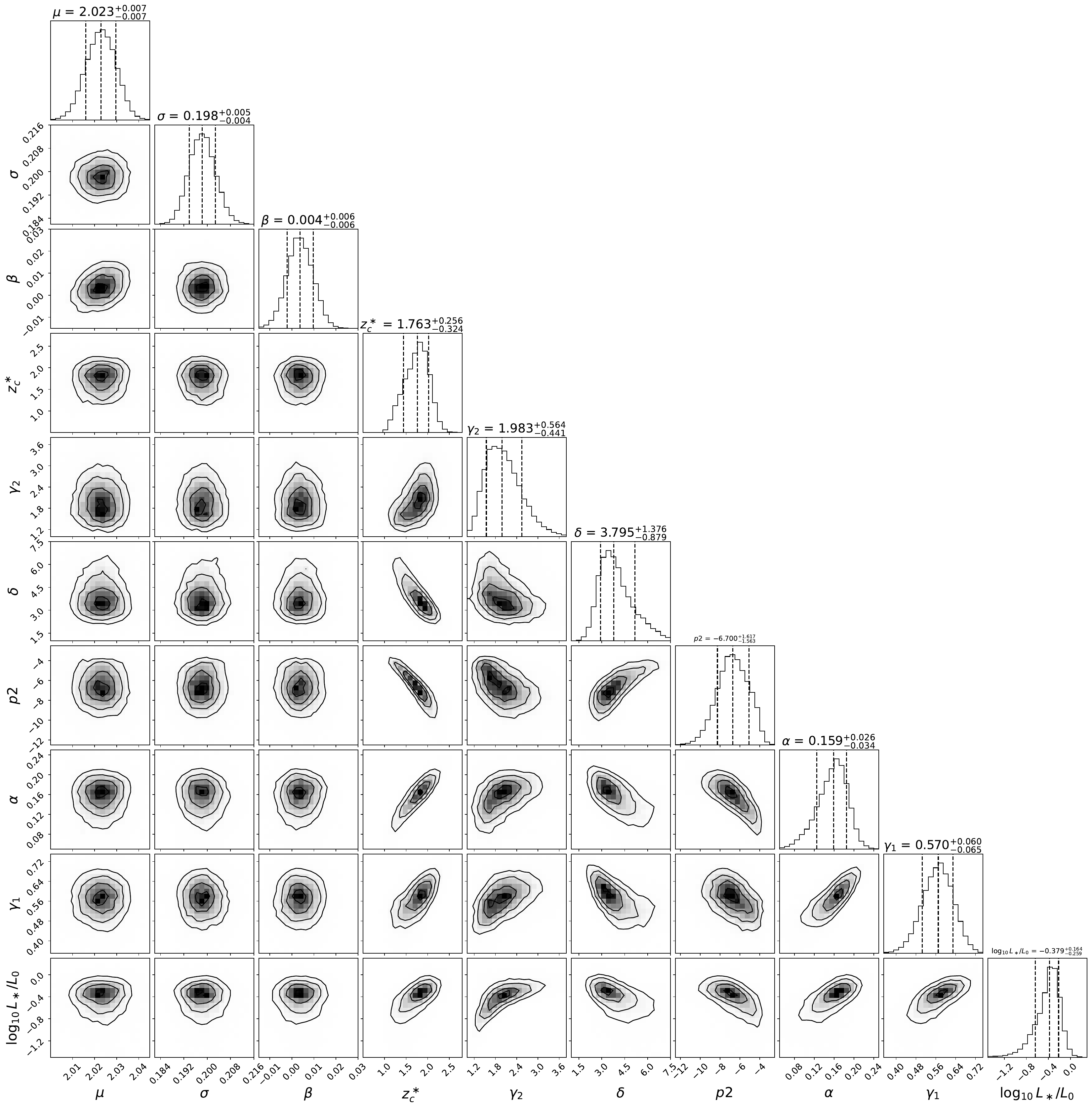}%
\caption{The marginalized posterior distribution and uncertainties of the model parameters obtained from the MCMC analysis of the LDDE model. The black dashed vertical lines on each histogram represent the 16th, 50th, and 84th percentiles, respectively.}
\label{fig:MCMC}
\end{figure*}

As an additional diagnostic, we performed an MCMC analysis of the LDDE model on one representative realization among the 1000 Monte Carlo realizations, and derived the best-fit parameters along with their uncertainties. The posterior distribution is sampled with 40 walkers, 5000 iterations, and 1000 burn-in using the python \texttt{emcee} package \citep{emcee_Foreman_2013}. 
Figure \ref{fig:MCMC} shows the corner plot originating from this analysis. We achieved a mean acceptance fraction of 0.36. We see that the parameter constraints are well behaved and unimodal. As expected from a complex model such as the LDDE, there are some parameter pairs that show correlation, such as $z_c^*$ and $\alpha$, which together determine the turnover redshift given the luminosity. Overall there are no strong boundary effects. This suggests that our maximum-likelihood optimizations are well localized. We also show in Table \ref{tab:LDDE_1realization} the best-fit parameters and uncertainties derived from our maximum-likelihood (ML) analysis, and compare them to the best-fit parameters of the MCMC analysis. The ML solution corresponds to the single combination of parameters that jointly maximizes the likelihood, whereas the MCMC values are medians of the one-dimensional marginalized posterior distributions. Correlations and asymmetries in the multidimensional posterior can therefore produce differences between these quantities. We see that the ML estimates are consistent with the MCMC results, with all parameter values lying within the corresponding 68\% intervals.

We note that Figure \ref{fig:MCMC} and Table \ref{tab:LDDE} characterize different sources of variation. Figure \ref{fig:MCMC} shows the posterior distributions of one Monte Carlo realization, whereas Table  \ref{tab:LDDE} reports the median and 68\% containment interval of the maximum-likelihood estimates obtained across all 1000 Monte Carlo realizations. The latter therefore is not expected to coincide exactly with the posterior distribution of any particular realization. For the realization shown in this section, however, we find the results are in agreement with the median parameters of Table \ref{tab:LDDE} within uncertainties.

\begin{deluxetable*}{lllllllllll}
\tablecaption{Comparison of best-fit parameters using the Maximum Likelihood (ML) and MCMC fit.}
\label{tab:LDDE_1realization}
\tablewidth{\textwidth}
\setlength{\tabcolsep}{3pt}
\tabletypesize{\scriptsize}
\tablehead{
\colhead{Method\textsuperscript{a}} & \colhead{$\mu$} & \colhead{$\sigma$} & \colhead{$\beta$} &
\colhead{$z^*_c$} & \colhead{$\ell_*$} & \colhead{$\gamma_1$} & \colhead{$\gamma_2$} &
\colhead{$\alpha$} & \colhead{$\delta$} & \colhead{$p2$}
}
\startdata
ML &
$2.022 \pm 0.006$ & 
$0.197 \pm 0.004$ &
$0.004 \pm 0.005$ &
$1.82 \pm 0.23$ &
$0.45 \pm 0.17$ &
$0.57 \pm 0.06$ &
$2.07 \pm 0.49$ &
$0.162 \pm 0.024$ &
$3.60 \pm 0.87$ &
$-7.15 \pm 1.43$ \\
MCMC &
$2.023^{+0.007}_{-0.007}$ &
$0.198^{+0.005}_{-0.004}$ &
$0.004^{+0.006}_{-0.006}$ &
$1.76^{+0.26}_{-0.33}$ &
$0.42^{+0.19}_{-0.19}$ &
$0.57^{+0.06}_{-0.06}$ &
$1.98^{+0.56}_{-0.44}$ &
$0.159^{+0.026}_{-0.034}$ &
$3.80^{+1.38}_{-0.88}$ &
$-6.70^{+1.62}_{-1.56}$
\enddata
\tablecomments{
\textsuperscript{a} Both methods are applied to the same Monte Carlo realization.
The ML best-fit parameters and uncertainties are obtained from the likelihood fit,
whereas the MCMC estimates represent the medians of the
one-dimensional marginalized posterior distributions,
with uncertainties corresponding to the 16th and 84th percentiles. \\
}
\end{deluxetable*}

\end{document}